\documentclass[12pt]{article}
\usepackage{titling}
\usepackage{amsmath}
\usepackage{slashed}
\usepackage{amssymb}
\usepackage{epsfig}
\usepackage{graphicx}
\usepackage{multirow,booktabs}
\usepackage{tabu}
\usepackage{color}
\usepackage[normal]{subfigure}
\usepackage{rotating}
\usepackage{hyperref,afterpage}
\usepackage[margin=1.0in]{geometry}
\usepackage[usenames,dvipsnames,svgnames,table]{xcolor}
\usepackage{enumitem}
\usepackage[utf8x]{inputenc}
\usepackage[compress,numbers,sort]{natbib}
\usepackage{authblk}
\usepackage{colortbl}
\usepackage{pdflscape}
\usepackage{braket}
\usepackage{color}
 \usepackage{pgfplotstable}
\usepackage{array}
\usepackage{colortbl}
\usepackage{float}
\usepackage{siunitx}
\usepackage{placeins}
\usepackage{comment}
\usepackage[compat=1.1.0]{tikz-feynman}
\usetikzlibrary{positioning,calc}

\tikzfeynmanset{warn luatex=false}

\newcommand{\RuleNode}[2]{%
  {$\displaystyle #2 $}%
}

\newcommand{\DiagWithRule}[2]{%
  \noindent
  \begin{minipage}[c]{0.58\textwidth}
    \centering
    #1
  \end{minipage}\hfill
  \begin{minipage}[c]{0.38\textwidth}
    \centering
    \RuleNode{0}{#2}
  \end{minipage}
  \par\bigskip
}
\definecolor{Gray}{gray}{0.95}

\usepackage{xparse}
\ExplSyntaxOn
\NewDocumentCommand{\longdash}{ O{2} }
{
	--\prg_replicate:nn { #1 - 1 } { \negthinspace -- }
}
\ExplSyntaxOff
\definecolor{nicered}{rgb}{0.6,0.1,0.1}
\definecolor{nicegreen}{rgb}{0.1,0.5,0.1}
\definecolor{mediumcandyapplered}{rgb}{0.99, 0.12, 0.07}
\definecolor{red}{rgb}{1.0, 0, 0}
\hypersetup{colorlinks,citecolor= nicegreen,linkcolor= nicered}

\def\det{\mbox{det}\,}

\renewcommand{\bar}{\overline}

\newcommand{\vs}{v_{\scriptscriptstyle S}}

\definecolor{LightCyan}{rgb}{0.88,1,1}
\definecolor{piggypink}{rgb}{0.99, 0.87, 0.9}
\definecolor{applegreen}{rgb}{0.55, 0.71, 0.0}
\definecolor{darkpastelgreen}{rgb}{0.01, 0.75, 0.24}
\definecolor{green-yellow}{rgb}{0.68, 1.0, 0.18}

\newcommand{\beq}{\begin{equation}}
\newcommand{\eeq}{\end{equation}}
\newcommand{\bea}{\begin{eqnarray}}
\newcommand{\eea}{\end{eqnarray}}

\newcommand{\published}[1]{%
\gdef\puB{#1}}
\newcommand{\puB}{}

\title{
\bf{Is the Standard Model Effective Field Theory Enough for Higgs Pair Production?}
}
\author[1]
{\'I\~nigo Asi\'ain\thanks{iasiain@icc.ub.edu}}
\affil[1]{\emph{\normalsize Departament de F\'isica Quantica i Astrof\'isica, Institut de Ci\`encies del Cosmos (ICCUB), Universitat de Barcelona, Mart\`i Franqu\`es 1, 08028 Barcelona, Spain}}
\author[2]
{Ramona Gr\"{o}ber\thanks{ramona.groeber@pd.infn.it}}
\affil[2]{\emph{\normalsize Dipartimento di Fisica e Astronomia ``G. Galilei", Universit\`a di Padova, and
Istituto Nazionale di Fisica Nucleare, Sezione di Padova,
I-35131 Padova, Italy}}
\author[3]
{Lorenzo Tiberi \thanks{lorenzo.tiberi@dottorandi.unipg.it}}
\affil[3]{\emph{\normalsize Dipartimento di Fisica e Geologia, Universit\`a di Perugia,  and
Istituto Nazionale di Fisica Nucleare, Sezione di Perugia, I-06123 Perugia,  Italy}}

\date{}

\published{\flushright 	COMETA-2026-05 \vskip1cm}
\begin{document} 

\maketitle

\begin{abstract}
\normalsize
We study Higgs-boson pair production in the Standard Model Effective Field Theory (SMEFT) up to dimension six and in the Higgs Effective Field Theory (HEFT) at leading order in the effective theory expansion, and assess which description is appropriate in concrete UV scenarios. Motivated by ``Loryon''-inspired models, we compare the Higgs pair production cross sections predicted by the full models to their SMEFT and HEFT counterparts. We identify regimes in which the two EFTs provide comparable descriptions, and clarify the limits required for their couplings to match. We also find that, for parts of parameter space in some of these models, HEFT can reproduce Higgs pair production more accurately than SMEFT, highlighting di-Higgs measurements as a potential probe of non-linear electroweak dynamics.
\end{abstract}

\clearpage

\section{Introduction}
After the discovery of the Higgs boson in 2012 \cite{ATLAS:2012yve, CMS:2012qbp}, huge process has been made in its characterisation \cite{ATLAS:2022vkf, CMS:2022dwd}. So far no significant deviation with respect to the Standard Model (SM) predictions has been found, neither by coupling measurements of the Higgs boson nor by direct searches for new physics. This motivates an effective field theory (EFT) approach that can be used to parameterise new physics effects in a model-independent way. 
\par
The two candidate EFTs are the Standard Model Effective Field Theory (SMEFT) \cite{BUCHMULLER1986621, dim6smeft, Contino:2013kra} and Higgs Effective Field Theory (HEFT) also known as electroweak chiral Lagrangian \cite{Feruglio:1992wf, Burgess:1999ha, Contino:2010mh, Buchalla:2012qq, Alonso:2012px,Brivio:2013pma,Buchalla:2015wfa}. They distinguish themselves on the assumptions made on the Higgs field as well as on the counting used for the ordering of the effective operators \cite{Gavela:2016bzc, Buchalla:2013eza,Brivio:2025yrr}. In SMEFT, the Higgs field transforms as an $SU(2)$ doublet as in the SM, while in HEFT the physical Higgs boson transforms as a singlet under the SM gauge group. In SMEFT the operators are ordered by their canonical dimension, whereas in HEFT a chiral dimension counting can be adopted \cite{Buchalla:2013eza, Brivio:2025yrr}.
\par
Recent years have seen substantial progress in clarifying when HEFT, rather than SMEFT, provides the appropriate low-energy description. More precisely, although the Higgs and Goldstone degrees of freedom can always be assembled into an 
$SU(2)$ doublet $H$, HEFT is required when the effective Lagrangian cannot be written as an analytic expansion around 
$H^{\dagger}H=0$, up to field redefinitions \cite{Falkowski:2019tft}.
In Refs.~\cite{Alonso:2015fsp, Alonso:2016btr, Alonso:2016oah, Cohen:2020xca} this criterion has been given a geometric interpretation: treating the Higgs and Goldstones as coordinates on a Riemannian scalar manifold, HEFT is needed when the scalar manifold has no fixed point of the 
$O(4)$ symmetry. A more phenomenological viewpoint was advocated in \cite{Gomez-Ambrosio:2022qsi, Gomez-Ambrosio:2022why}, where the necessity of HEFT over SMEFT was linked to the presence of zeros in the flare function.
\par 
Clearly, HEFT is more general than SMEFT \cite{Brivio:2017vri} and generically predicts larger deviations with a validity of the framework up to a scale of $4\pi v$ where $v$ is the vacuum expectation value of the scalar field. While in the SMEFT the SM limit can be achieved smoothly by setting the scale that suppresses the operators of higher dimensions to large values, in the HEFT the SM limit is obtained by tuning the coefficients in front of the operators to one or zero. 
\par 
In this paper, we will study from a phenomenological point of view whether it will be necessary to adopt the more general HEFT for studies in Higgs pair production or whether it is sufficient to describe potential deviations in terms of SMEFT by adopting concrete UV scenarios and connecting them to HEFT and SMEFT. 
\par
While Higgs pair production is usually considered a probe of the trilinear Higgs self-coupling \cite{Djouadi:1999rca, Dolan:2012rv, Baglio:2012np}, it  can also be used to probe \emph{Higgs non-linearities} \cite{Grober:2010yv, Contino:2012xk, Grober:2016wmf} hence whether the couplings of one or several Higgs bosons are connect via the linear EFT relations given in SMEFT. In particular in Higgs pair production in vector boson fusion, one can directly probe the \textit{flare} function that distinguishes between HEFT and SMEFT in the two-derivative case \cite{Gomez-Ambrosio:2022qsi, Gomez-Ambrosio:2022why, Delgado:2023ynh}. This is due to the fact that in SMEFT couplings of one Higgs boson to vector bosons, gluons and fermions are usually correlated with the equivalent coupling to two Higgs bosons at the same order in the EFT expansion, while in the HEFT they are not correlated.
We notice though that, as has been shown recently in \cite{Grober:2025vse} for Higgs pair production in gluon fusion, that this is purely a question about which EFT converges faster. For this reason, we stay at the order in the EFTs commonly adopted in current experimental analyses -- LO HEFT Lagrangian and dimension six in SMEFT.

The couplings of two Higgs boson to fermions or vector bosons are already probed by the ATLAS and CMS experiments \cite{CMS:2022gjd,ATLAS:2022fxe}.
One of the questions we assessed in this work, is whether the current bounds on these kind of couplings probe realistic parameter space in terms of concrete models.

In a first step, we identify candidate UV scenarios in which the appropriate low-energy description is HEFT rather than SMEFT. A well-defined class of such models was introduced in Ref.~\cite{Banta:2021dek} under the name Loryons, building on the criterion of Ref.~\cite{Cohen:2020xca}: the new states should acquire more than half of their mass from electroweak symmetry breaking. When this condition is met, the SMEFT expansion is not the natural organizing principle and a HEFT description is required. Phenomenological studies of Loryons include, for instance, Refs.~\cite{Crawford:2024nun, Kilic:2026ogm}. In our analysis we select representative Loryon setups and augment their Lagrangians with additional interactions that generate tree-level matching coefficients, thereby enhancing their impact on Higgs pair production. More generally, HEFT can also be necessitated by scenarios with additional sources of electroweak symmetry breaking \cite{Cohen:2020xca}.

For three benchmark models -- the singlet extension of the SM, the two-Higgs-doublet model, and a colored-scalar extension -- we compute their contributions to Higgs pair production in the UV theory and compare them to the corresponding descriptions in SMEFT and HEFT. This allows us to delineate in a concrete way when HEFT is required, and when SMEFT remains sufficient, for interpreting Higgs-pair searches. Related questions have been explored for Higgs pair production in vector boson fusion in Ref.~\cite{Englert:2023uug}.

The paper is structured as follows. In section \ref{sec:EFT} we discuss the two different EFT parameterisations of Higgs pair production. In section \ref{sec:UV} we first discuss how we identify the considered models and then discuss them one by one. We offer our conclusions in section \ref{sec:conclusion}.

%%%%%%%%%%%%%%%%%%%%%%%%%%%%%%%%%%%%%%%%%%%%%%%%%%%%%%
\section{Effective Field Theory Description of Higgs pair production \label{sec:EFT}}
A Higgs boson pair is dominantly produced in the SM by gluon fusion. The cross section is around 37 fb at $\sqrt{s}=14\text{ TeV}$ \cite{Dawson:1998py, Baglio:2018lrj, Borowka:2016ehy, Borowka:2016ypz, Grazzini:2018bsd, AH:2022elh, Bagnaschi:2023rbx}. The second most important Higgs pair production process is vector boson fusion (VBF) with a cross section of 2 fb at $\sqrt{s}=14\text{ TeV}$ \cite{Baglio:2012np, Ling:2014sne, Dreyer:2018qbw, Dreyer:2020xaj}.
In effective field theory descriptions those two processes receive modified couplings as well as new structures from effective operators. We define the SM Lagrangian as
 \begin{align}
   \mathcal{L}&=(D_{\mu} H)^{\dagger} (D^{\mu}H)- V(H)  - \frac{1}{4}B_{\mu\nu}B^{\mu\nu} - \frac{1}{4}W_{\mu\nu}^a W^{\mu\nu,a} -\frac{1}{4}G_{\mu\nu}^aG^{\mu\nu,a} \nonumber\\&-\left(y_d \bar{q}_L H d_R+y_u  \bar{q}_{L} \tilde{H} u_R 
  + y_e \bar{\ell}_L H e_R +\text{h.c}\right)
+\sum_{\psi=q_L,\ell_L, u_R, d_R, e_R}i\bar{\psi}\slashed{D}\psi\, ,
 \end{align}
 with $\tilde{H}_i=\epsilon_{ij}H_j$, the $\ell_L$ and $q_L$ are the $SU(2)_L$ lepton and quark doublets, $e_R, d_R, u_R$ the $SU(2)_L$ singlets, $B_{\mu\nu}, W_{\mu\nu}, G_{\mu\nu}$ the $U(1)$, $SU(2)$ and $SU(3)$ field strenghts and the Higgs potential is given by 
 \begin{equation}
 V(H)=\mu_1^2 |H|^2+\lambda_H |H|^4\,,
 \end{equation}
 with $H$ the scalar doublet field, taking the form $H=1/\sqrt{2}(0,v_H + h)^T$ in the unitary gauge.
 For our considerations the following SMEFT operators are relevant (using the so-called Warsaw basis \cite{dim6smeft})
\begin{equation}
\begin{split}
\Delta\mathcal{L}_{\text{SMEFT}}&=\frac{C_{H,\Box}}{\Lambda^2} (H^{\dagger} H)\Box (H^{\dagger } H)+ \frac{C_{H D}}{\Lambda^2}(H^{\dagger} D_{\mu}H)^*(H^{\dagger}D^{\mu}H)+ \frac{C_H}{\Lambda^2} (H^{\dagger}H)^3 \\ &+\left( \frac{C_{tH}}{\Lambda^2} H^{\dagger}{H}\bar{Q}_L\tilde{H}\, t_R + h.c.\right)+\frac{C_{H G}}{\Lambda^2} H^{\dagger} H G_{\mu\nu}^a G^{\mu\nu,a}\,, \label{eq:Warsaw}
%\\ &+\frac{C_{uG}}{\Lambda^2} 
%(\bar Q_L\sigma^{\mu\nu}T^aG_{\mu\nu}^a\tilde{H} \,t_R +{\rm h.c.})\,, \label{eq:Warsaw}
\end{split}
\end{equation}
where $Q_L$ stands for the third generation quark doublet.
The operators $\mathcal{O}_{H,\Box}$, $\mathcal{O}_{HD}$ and $\mathcal{O}_{H}$ modify Higgs pair production via gluon fusion and vector boson fusion, while all the operators of Eq.~\eqref{eq:Warsaw} affect Higgs pair production in gluon fusion. We do not explicitly write down operators that affect the light quark couplings to the Higgs boson as their importance depends on the flavour structure assumed, note though that they can be also be potentially probed in Higgs pair production \cite{Alasfar:2019pmn, Alasfar:2022vqw,Celada:2023oji, Han:2023njx}. We assume CP-conservation and refer to \cite{Grober:2017gut} for a study of CP-violating couplings in Higgs pair production. Furthermore, we omit operators of type $H^{\dagger} H W_{\mu\nu}W^{\mu\nu}$ or ($\bar{Q}_L \sigma^{\mu\nu}T^A t_R) \tilde{H}G^A_{\mu\nu}$ adopting a loop counting \cite{Arzt:1994gp, Buchalla:2022vjp}. The chromomagnetic operator enters into one-loop diagrams but is loop-generated, hence counts effectively as a two-loop contribution. When adopting the loop counting it  enters consistently at the same order as e.g.~four-top operators \cite{DiNoi:2023ygk,Heinrich:2023rsd}. The operator $\mathcal{O}_{HG}$ is also loop-generated but entering in tree-level diagrams in gluon fusion respects the considered order. As was pointed out in Ref.~\cite{Alasfar:2023xpc}, it is useful to extract from the operator $\mathcal{O}_{HG}$ a factor $\alpha_s(\mu)$
\begin{equation}
C_{HG} \to \tilde{C}_{HG}=\frac{1}{\alpha_s(\mu)} C_{HG}\,,
\end{equation}
since $\alpha_s$ is usually evaluated at a dynamical scale in Higgs pair production via gluon fusion.

The Warsaw basis is constructed such that derivative operators are systematically removed by equations of motion, but two derivative Higgs interactions remain. Since they contain covariant derivatives 
they cannot be removed by gauge-independent field redefinitions. In order to obtain a canonically normalised Higgs kinetic term, the standard field redefinition (in unitary gauge) is 
\begin{equation}
H=\frac{1}{\sqrt{2}}\left( \begin{array}{c} 0 \\ h(1+v^2\frac{C_{H,\textrm{kin}}}{\Lambda^2}) + v \end{array} \right)\label{eq:fieldredefinition}
\end{equation} 
with 
\begin{equation}
C_{H,\textrm{kin}}=\left(C_{H,\Box}-\frac{1}{4}C_{HD}\right)\,.
\end{equation}
In Higgs pair production it turns out to often be convienient to adopt yet another field redefinition which removes derivative Higgs self-interactions
\begin{equation}
h \to h + v^2\frac{C_{H,\textrm{kin}}}{\Lambda^2}\left( h +\frac{h^2}{v}+\frac{h^3}{3v^2}\right)\,. \label{fieldref}
\end{equation}
In this way, the Higgs trilinear self-coupling gets modified by a multiplicative factor only with respect to the SM and no new derivative structures come into play. On the other hand, it leads to a dependence on $C_{H,\textrm{kin}}$ for all Higgs boson couplings. In the following, we will use the field redefinition of Eq.~\eqref{fieldref}.
\par
We turn now to discuss HEFT. In that case, there is no longer a Higgs doublet structure $H$  but the Goldstone bosons $\pi^i$, which are associated to the breaking of the $SU(2)_L\times U(1)_Y \to U(1)_Q$ can be described by the matrix 
\begin{equation}
\Sigma=e^{i\sigma^i \pi^i/v} \hspace*{0.5cm}\text{with } \hspace*{0.5cm}v=246\text{ GeV}\, \hspace*{0.5cm}\text{and } \hspace*{0.5cm}i=1,2,3 .
\end{equation}
The physical Higgs field $h$ arises as a singlet. The relevant Lagrangian for our purpose is given by
\begin{align}
\Delta{\cal L}_{\text{HEFT}} &=\frac{v^2}{4}\text{Tr}\left(D_{\mu}\Sigma^{\dagger} D^{\mu} \Sigma\right)\left(1+2 c_{hVV}\frac{h}{v}+c_{hhVV}\frac{h^2}{v^2} \right)
-m_t\left(c_t\frac{h}{v}+c_{2t}\frac{h^2}{v^2}\right)\,\bar{t}\,t \\ & -
c_{hhh} \frac{m_h^2}{2v} h^3+\frac{\alpha_s}{8\pi} \left( c_{ggh} \frac{h}{v}+
c_{gghh}\frac{h^2}{v^2}  \right)\, G^a_{\mu \nu} G^{a,\mu \nu}\;. \nonumber \label{eq:ewchl}
\end{align}
We note that the operators with coefficients $c_{ggh}$ and $c_{gghh}$ contribute an order higher in the chiral counting\footnote{Counting also the orders of $g_s$ into the chiral dimension \cite{Brivio:2025yrr}.} with respect to the other terms, but they arise at the same order once inserted into Feynman diagrams since they generate tree-level diagrams \cite{Brivio:2025sib}. 
\begin{table}[ht]
\centering

\begin{tabular}{|c|c|}
   \hline
                  HEFT       &   SMEFT                 \\
    \hline
           &      \\
   $c_{hVV}$    & \quad $1+ \frac{C_{H,\textrm{kin}}v^2}{\Lambda^2}$ \\
          &      \\
   \hline 
           &      \\
   $c_{hhVV}$  &    \quad $1+ \frac{4C_{H,\textrm{kin}}v^2}{\Lambda^2}$ \\
            &      \\
   \hline
          &      \\
   $c_{hhh}$  & \quad  $1 - \frac{2 v^2}{m_{h}^2} \frac{v^2 C_{H}}{\Lambda^2} + 3 \frac{v^2}{\Lambda^2} C_{H,\textrm{kin}}$ \\
        &      \\
   \hline
         &      \\
   $c_{t} $    &  \quad  $1 - \frac{v^2}{\sqrt{2} \Lambda^2}  \frac{v}{m_t} C_{uH} + \frac{v^2}{\Lambda^2} C_{H,\textrm{kin}}$ \\
             &     \\
   \hline
             &      \\
   $c_{2t}$   &  \quad  $- \frac{3 v^2}{2 \sqrt{2} \Lambda^2}  \frac{v}{m_t} C_{uH} + \frac{v^2}{\Lambda^2} C_{H,\textrm{kin}}$ \\
            &          \\
    \hline
             &       \\
    $c_{ggh} $    &  \quad  $\frac{8\pi}{\alpha_s} \frac{v^2}{\Lambda^2} C_{HG}$ \\
             &     \\
   \hline
             &      \\
   $c_{gghh}$   &  \quad  $\frac{4\pi}{\alpha_s} \frac{v^2}{\Lambda^2} C_{HG}$ \\
       &          \\
   \hline
\end{tabular}
\caption{Relevant couplings in di-Higgs production.}
\label{tab::di-Higgs}
\end{table}
The relevant couplings for Higgs pair production are given in Table~\ref{tab::di-Higgs}.
We can see that HEFT provides the more general description than SMEFT. In SMEFT at dimension 6, the couplings of two Higgs boson and one Higgs boson to bosons or fermions are correlated, while they are not in HEFT. There, the couplings to two or more Higgs boson arise at the same order as the coupling of one Higgs boson as the Higgs boson does not carry chiral dimension. Of course also in SMEFT the couplings of one or two Higgs bosons to bosons or fermions can become decorrelated, but this requires to go to operators of higher dimensions. For instance at dimension 8 an operator $|H|^4G_{\mu\nu}G^{\mu\nu}$ is part of the dimension-8 Lagrangian, decorrelating the one Higgs to gluon and the two Higgs to gluon coupling. In this case, there is however a suppression of the decorrelation effect by $1/\Lambda^2$. We refer to \cite{ggHHdim8} for a study of dimension-8 effects in Higgs pair production in gluon fusion and to \cite{Dedes:2025oda} in vector boson fusion. Higher order effects in HEFT have been considered in \cite{Herrero:2022krh} for vector boson fusion and we refer to \cite{Brivio:2025sib} for gluon fusion HEFT Higgs pair production including also operators beyond the LO Lagrangian.

%%%%%%%%%%%%%%%%%%%%%%%%%%%%%%%%%%%%%%%%%%%%%%%%%%%%%
\section{UV Models \label{sec:UV}}
In order to answer the question whether it is sufficient to adopt SMEFT in Higgs pair production or whether HEFT should be used, we study three UV models, which are inspired by the \textit{Loryon} catalogue \cite{Banta:2021dek}. We chose as examples a model that generates via tree-level matching contributions to $C_{H,kin}$ and hence a rescaling to all Higgs couplings -- the singlet scalar model. With respect to \cite{Banta:2021dek} we do not impose a $\mathbb{Z}_2$ symmetry to allow for larger effects. For this case we study both Higgs pair production in gluon fusion and vector boson fusion.
Then, we move to a model where the main effect can come from Higgs couplings to top quarks -- the two Higgs doublet model (2HDM). The latter has been studied in the context of \textit{Loryons} in \cite{Kilic:2026ogm}. We focus our study to gluon fusion for this model. What regards vector boson fusion, we refer to the study in \cite{Arco:2023sac}.  Finally, we discuss a model whose leading effects are in the Higgs gluon couplings -- a  model featuring colored scalars. 
%\begin{itemize}
%    \item connection geometry, flare function
%    SMEFT and HEFT show their difference if we take a geometric approach, namely identifying the scalar sector of our EFT as a scalar manifold where fields are now coordinates \cite{Alonso:2016oah}. A more phenomenological approach is pursued through the Flare function \cite{Gomez-Ambrosio:2022qsi}, identifying the values for which it vanishes as a necessary condition for the consistency of passing from HEFT to SMEFT.
%    \item connection potential, derivative terms
%    \item connection to the mass criteria
%    \item extend the loryons with the explicit terms, use as criteria that LO SMEFT in our couplings is generated
%\end{itemize}
%%%%%%%%%%%%%%%%%%%%%%%%%%%%%%%%%%%%%%%%%%%%%%%%%%%%%
\subsection{Scalar Singlet Model \label{sec:singlet}}
We started with the scalar singlet model. The SM is augmented with an additional real scalar field $\Phi$. The potential then takes the following form
\begin{equation}
V(H,\Phi) = 
\mu_1^2 |H|^2+\lambda_H |H|^4+\frac{1}{2} \mu_2^2\Phi^2+\mu_4 |H|^2 \Phi +\frac{1}{2}\lambda_3 |H|^2 \Phi^2+\frac{1}{3} \mu_3 \Phi^3+\frac{1}{4}\lambda_2 \Phi^4 \,.\label{eq:singlet}
\end{equation}
A tadpole term for the singlet field has been omitted, as it can be reabsorbed in the singlet vacuum expectation value by a field redefinition.

The term with the $\mu_4$ coupling unavoidably induces a vacuum expectation value for $\Phi$ and also leads to a mixing between $H$ and $\Phi$. On the other hand it also leads to tree-level generated effective operators for the singlet model. The potential becomes $\mathbb{Z}_2$ symmetric if $\mu_4=\mu_3=0$. We consider here though the more general case where they are free parameters.

The scalar fields can be expanded around their vacuum expectation values by
\begin{equation}
H=\frac{1}{\sqrt{2}} \left( \begin{array}{c} 0 \\ v_H +h \end{array} \right) \, ,  
\qquad 
\Phi= (v_S+S)
\,, 
\end{equation}
employing the unitary gauge for the Higgs doublet. 

The tadpole conditions are given by
\begin{align}
-\mu_4 v_S-\frac{\lambda_3 v_S^2}{2}-\mu_1^2-\lambda_H v_H^2= 0 \,,\\
 -\frac{\mu_4 v_H^2}{2 }-\frac{1}{2} \lambda_3 v_H^2 \vs-\mu_2^2 v_S-\lambda_{2}  v_S^3-\mu_3  v_S^2= 0 \,. \label{eq:tadpolecond}
\end{align}
With the first condition one can replace $\mu_1^2$ in terms of $v_H$.
The mass terms can be diagonalised by rotation with a matrix \begin{equation}
\left(\begin{array}{c} h_1 \\ h_2 \end{array}\right)=\left( \begin{array}{cc} \cos\theta & \sin\theta \\ -\sin\theta & \cos\theta \end{array}\right) \left(\begin{array}{c} h \\ S \end{array}\right)\,,
\label{eq:mixingangle}
\end{equation}
such that the mass matrix $M^2$ becomes diagonal.

The mass matrix in the real $(h,S)$ basis then reads
\begin{equation}
M^2 =
\left(\begin{array}{cc} m_{hh} & m_{hS} \\
m_{hS} & m_{SS}
 \end{array}\right) \, ,
 %\left( \begin{array}{c} h \\ S \end{array}\right)
\end{equation}
with
\begin{align}
m_{hh}&= 2 v_H^2 \lambda_H \,,\\
m_{hS}&=v_H \left(\mu_4 + \lambda_3 v_S \right)\,,\\
m_{SS}&= \mu_2^2 +\frac{1}{2} \left(\lambda_3 v_H^2 +6 v_S^2 \lambda_{2}+4 v_S \mu_3 \right)\,,
\end{align}
and the mass eigenvalues are
\begin{align}
m_{1,2}^2&=\frac{1}{2}\left(m_{hh}+m_{SS} \mp \sqrt{4 m_{hS}^2 +(m_{hh}-m_{SS})^2}\right) \nonumber \\
&= \frac{1}{2}\left( m_{hh}+ m_{SS} \pm (m_{hh}-m_{SS}) \frac{1}{\cos 2 \theta}\right) \, .
\end{align}
The mixing angle $\theta$ is given by the relation
\beq 
\tan 2 \theta =\frac{2 m_{hS}}{m_{SS}-m_{hh}} \, .
\eeq 
We can treat now some of the potential parameters in terms of the masses $m_1$ and $m_2$, the vacuum expectation values $v_H$ and $v_S$ as well as the mixing angle $\theta$ via the following relations
\begin{align}
\mu_1^2 &=- \frac{1}{4}\left[\left(-2 \lambda_3 v_S^2+m_1^2+m_2^2 \right)+ \cos 2 \theta \, (m_1^2-m_2^2)
- 2 \frac{v_S}{v_H}\sin 2 \theta \,
   (m_1^2-m_2^2) \right]\,,  \label{eq:mu1}\\
%%%
\mu_2^2 &=\frac{1}{2}\left[\left( \lambda_3 v_H^2- m_1^2- m_2^2+2\lambda_{2}
   v_S^2\right)+\frac{v_H}{v_S} \sin 2\theta \, (m_1^2-m_2^2)+ \cos 2\theta \,(m_1^2-m_2^2)\right]\,,\label{eq:mu2} \\
%%%   
   \mu_3 &= \frac{1} {2 v_S} \left( m_1^2+
   m_2^2- \lambda_3 v_H^2 -4\lambda_{2}  v_S^2\right)-\frac{1}{4} \frac{v_H}{v_S^2} \sin 2\theta\,
   (m_1^2-m_2^2)  \nonumber \\  & - \frac{1} {2 v_S} \cos 2\theta\,
   (m_1^2-m_2^2)\,, \label{eq:mu3}\\
   %%%%%%%%
   \mu_4& = \frac{\sin 2 \theta\, (m_2^2-m_1^2)-2 \lambda_3 v_H v_S}{2
   v_H}  \,, \label{eq:mu4}\\
   \lambda_H &= \frac{\cos 2\theta \, (m_1^2-m_2^2)+m_1^2+m_2^2}{4
   v_H^2}\,.  \label{eq:lambdah}
\end{align}
%It is convenient to write the broken phase UV lagrangian in terms of the physical fields $h_1$ and $h_2$:
%\begin{align}
%  \label{eq::HEFT Scalar singlet}
% \mathcal{L}_{UV} =  & -\frac{1}{4} G_{\mu \nu}^{a} G^{a \mu \nu} - \frac{1}{4} W_{\mu \nu}^{I} W^{I \mu \nu} - \frac{1}{4} B_{\mu \nu} B^{\mu \nu} 
%- V(h_1, h_2)  +\frac{v^2}{4} \operatorname{Tr}\left[(D_{\mu} \Sigma)^\dag D^{\mu} \Sigma \right] \notag \\
% & \Bigl(1 + 2 \cos(\theta) \frac{h_1}{v} + 2 \sin(\theta) \frac{h_2}{v}+ \cos(\theta)^2 \Bigl(\frac{h_1}{v}\Bigr)^2 + \sin(\theta)^2 \Bigl(\frac{h_2}{v}\Bigr)^2 + 2 \cos(\theta) \sin(\theta) \frac{h_2 h_1}{v^2} \Bigr)  \notag \\
%+ & \frac{1}{2} (\partial_{\mu} h_1)(\partial^{\mu}h_1) + \frac{1}{2} (\partial_{\mu} h_2)(\partial^{\mu} h_2) + i \sum_{j = l, q, e, \nu, d, u} \bar{\Psi}_j \slashed{D} \Psi_{j} +\notag \\
%-  & \frac{v}{\sqrt{2}} \left( \bar{q}_{L} \Sigma Y_{Q} q_{R} + Y_{L} \bar{l}_{L} \Sigma Y_{L} l_{R} + \text{h.c.}\right) 
%\Bigl(1 + \cos(\theta) \frac{h_1}{v} + \sin(\theta) \frac{h_2}{v}\Bigr) ,
%\end{align}
%where $Y_Q = \text{diag}(Y_u , Y_d)$, $Y_l = \text{diag}(0, Y_e) $ and the right handed doublets for single generation are given by: 
%\begin{equation}
%    q_{R}= \begin{bmatrix}
%        u_R \\
%        d_R\\
%    \end{bmatrix} ,  \quad l_{R} = \begin{bmatrix}
%        0 \\
%        e_R \\
%    \end{bmatrix} \,.
%\end{equation}
The scalar potential in the broken phase can be written as
\begin{align}
    V(h_1, h_2) =  & \frac{m^2}{2} h_{1}^2 + \frac{ M^2}{2} h_{2}^2+d_{1} h_{1}^3 +d_2 h_{1}^2 h_{2}+d_3 h_1 h_{2}^2 + d_4 h_{2}^3 + z_1 h_{1}^4  \notag \\
    + & z_2 h_{1}^3 h_{2} + z_3 h_{1}^2 h_{2}^2 + z_4 h_{1} h_{2}^3 + z_5 h_{2}^4 , \label{eq:singletpotential}
\end{align}
where $d_i$,$z_i$ $i=1,..,5$ are expressed in terms of UV model parameters. The explicit expressions for coefficients relevant for Higgs pair production can be found in Appendix~\ref{app:couplings}.

\subsubsection{Constraints on the parameter space \label{subsec:singlet_parameterscan}}
The scalar singlet model has a relatively vast parameter space, which we restrict by using theoretical and experimental constraints.
\begin{description}
\item[Vacuum stability:] 
We demand that the potential $V(H, \Phi)$ is bounded from below, and we additionally impose that the electroweak vacuum corresponds to a (local) minimum. For large field values the relevant terms are
\begin{equation}
V(H,\Phi)\simeq \lambda_H\,(H^\dagger H)^2 + \frac{\lambda_2}{4}\,\Phi^4 + \frac{\lambda_3}{2}\,(H^\dagger H)\Phi^2 \, .
\end{equation}
Introducing $x\equiv H^\dagger H\ge 0$ and $y\equiv \Phi^2\ge 0$, one may rewrite
\begin{equation}
V \simeq \left(\sqrt{\lambda_H}\,x - \frac{\sqrt{\lambda_2}}{2}\,y\right)^2
+ \left(\frac{\lambda_3}{2}+\sqrt{\lambda_H\lambda_2}\right) x\,y \, .
\end{equation}
It follows that the potential is bounded from below if
\begin{equation}
\lambda_H(\mu)>0,\qquad \lambda_2(\mu)>0,
\end{equation}
and, in addition, for $\lambda_3(\mu)<0$,
\begin{equation}
\lambda_3(\mu) > -2\sqrt{\lambda_H(\mu)\lambda_2(\mu)}
\qquad \Longleftrightarrow \qquad
4\,\lambda_H(\mu)\lambda_2(\mu) - \lambda_3(\mu)^2 >0 \, .
\end{equation}

%We impose these conditions on the running couplings up to a maximum scale $\Lambda$ \RG{comment which $\Lambda$?} using the one-loop renormalization-group evolution.

To ensure that the electroweak vacuum is a local minimum, we also require the scalar mass matrix (the Hessian of $V$ evaluated at the vacuum configuration) to be positive definite. In particular, a necessary condition is that its determinant is positive:
\begin{equation}
\det M^2 > 0 \, ,
\end{equation}
which in our parametrization leads to
\begin{align}
\left(2 \lambda_H v_{H}^2 \right) 
\left( 2 \lambda_2 v_{S}^{2} + \mu_3 v_S - \frac{ v_{H}^2 \mu_4}{2 v_S} \right) 
> \left( \lambda_{3}^2 v_{H}^2 v_{S}^2 + \mu_{4}^2 v_{H}^2 + 2 \lambda_3 \mu_4 v_{H}^2 v_{S} \right) \,.
\end{align}
\item[Perturbativity:]
We require perturbativity of the couplings from the electroweak scale up to Planck scale employing one-loop RGEs. This entails a generic upper bound on $|\lambda_i (\Lambda)| \leq 4\pi$, such that strongly coupled regime is avoided. \\
%\RG{THis point should go together with perturbative unitarity.}
For coefficients with mass dimension one we require that the one-loop corrections to the trilinear Higgs self coupling are smaller than their tree-level values in the $SU(2)$ limit \cite{DiLuzio:2017tfn}. This leads to:
\begin{equation}
    \frac{|\mu_4|}{\text{max} \left(|\mu_2|, |\mu_1|\right)} \leq 4 \pi ,\quad \land \quad \left|\frac{\mu_3}{\mu_2} \right| \leq 4\pi \,.
\end{equation}
%In the SM $\lambda_1$ becomes negative at an energy scale $\Lambda_{\text{inst}}\sim 10^9$GeV, however new $1$-loop corrections due to the scalar singlet keeps $\lambda_1$ positive up to the Planck scale, this result is highly sensitive to the $\lambda_3$ value at the EW scale.
%\RG{Do we impose anything there or is this a general point?} \LT{It's a just a general point}.
Bounds on the couplings of the model can be imposed by exploiting perturbative unitarity \cite{PhysRevD.16.1519}.
We compute the partial wave amplitudes $a_{i,j \rightarrow kl}^{0}$ for $J=0$ for every possible $2 \rightarrow 2$ scattering amplitude with $i,j,k,l$ either $h$ or $S$. We check that the model does not violate $|\text{Re} (a_{i j \rightarrow k l}^{0})| < 1/2$, for each eigenvalue of the scattering matrix.
In the vicinity of the poles the scattering amplitudes diverge as an indication that higher order corrections are needed. Since we are neglecting the width effects in this tree level analysis, we follow \cite{Goodsell:2018tti} and apply a kinematic cut of
\begin{equation}
    \left| 1- \frac{s}{m^2}\right| > 0.25 ,
\end{equation}
where $s$ denotes the Mandelstam variable for $2 \rightarrow 2$ and $m$ is the Higgs mass.

\item[Electroweak precision tests:] As reported in \cite{Lopez-Val:2014jva} the strongest bound on the mixing angle comes from measurements of the $W$ boson mass, $M_{W}$. Using the results of Ref.~\cite{Papaefstathiou:2022oyi} a bound of $|\sin(\theta)|<0.155$ holds for physical scalar singlet masses $m_2 >800$ GeV. This bound is not updated to incorporate the more recent results of the ATLAS and CMS collaborations \cite{ATLAS:2023fsi, 2024mWCMS} which further shrink the uncertainty, but no official combination exists so far.

\item[Higgs coupling modifiers:] Recent measurements from the ATLAS and CMS collaboration have fixed possible deviations for Higgs coupling to massive gauge bosons and to fermions in the framework of $\kappa$ formalism \cite{ATLAS:2022vkf, CMS:2022dwd}. At $2 \sigma$ confidence we have $\sin \theta \leq 0.3741$. This bound results less strong than the one of the $M_{W}$ measurement.

\end{description}
\subsubsection{Effective Field Theory for the singlet model}
First we will discuss the contributions of the singlet model to SMEFT operators that modify the couplings entering the Higgs pair production process. As can be inferred from Refs.~\cite{deBlas:2017xtg, Jiang:2018pbd, Haisch:2020ahr}, the singlet model generates at tree-level the effective operators $\mathcal{O}_{H \Box}$ and $\mathcal{O}_H$. In addition, it generates a contribution to $|H|^4$ that shifts $\lambda_H$ to \[ \lambda_H \xrightarrow{} \lambda_H - \frac{\mu_{4}^2}{ \mu_{2}^2}  \,. \]
The other Wilson coefficients are:
\begin{align}
  & \frac{C_{H\Box}}{\Lambda^2} = -\frac{\mu_{4}^2}{2 \mu_{2}^4}  ,  \\
  & \frac{C_{H}}{\Lambda^2} = - \frac{\lambda_3 \mu_{4}^2}{2 \mu_{2}^4} + \frac{\mu_3 \mu_{4}^3}{3 \mu_{2}^6}   \,.   
\end{align}
On-shell matching to the HEFT basis has been performed through a diagrammatic approach. We obtain:
%\\
%A non-linear relation between $c_{hVV}$ ($c_{t}$) and $c_{hhVV}$($c_{2t}$) is a consequence of HEFT being more general than SMEFT where the same coefficients are linearly related:
%\begin{align}
%      & \mathcal{M}_{\text{HEFT}} (V V\rightarrow h) =  \mathcal{M}_{\text{UV}} (V V\rightarrow h)    \rightarrow & c_{hVV} &= \cos(\theta) , \notag \\            
 %      & \mathcal{M}_{\text{HEFT}} (V V \rightarrow h h) = \mathcal{M}_{\text{UV}} (V V \rightarrow h h)     \rightarrow & c_{hhVV} &=\cos(\theta)^2 - 2 \frac{\sin(\theta) v d_2}{M^2}, \notag \\     
%        & \mathcal{M}_{\text{HEFT}} (h \rightarrow h h) =\mathcal{M}_{\text{UV}} (h \rightarrow h h)      \xrightarrow{} &c_{hhh} & = \frac{2v}{m^2} d_1 , \notag \\
%        & \mathcal{M}_{\text{HEFT}} (t\Bar{t} \rightarrow h) =  \mathcal{M}_{\text{UV}} (t\Bar{t} \rightarrow h)   \xrightarrow{} &c_{t} &=\cos(\theta) ,\notag \\
 %        & \mathcal{M}_{\text{HEFT}} (t \Bar{t} \rightarrow h h) =  \mathcal{M}_{\text{UV}} (t \Bar{t} \rightarrow h h) \xrightarrow{}  &c_{2t} & =-  \frac{\sin(\theta) v d_2}{M^2}  \,. \notag  \\        
    \label{tab:my_label}
%\end{align}
\begin{align}
&c_{hVV} = \cos \theta\,, \quad c_{hhVV} =\cos^2\theta - 2 \frac{\sin \theta\, v d_2}{m_2^2}\,, \quad c_{hhh}  = \frac{2v}{m_1^2} d_1\,, \\ & c_{t} =\cos\theta \,, \quad c_{2t}  =-  \frac{\sin \theta\, v d_2}{m_2^2}\,. \nonumber
\end{align}
%The SM limit is obtained setting $c_{hVV}=1, c_{hhVV}=1, c_{hhh}=1, c_{t}=1, c_{2t}=0$.
%The s Mandelstam variable is assumed to be $s \ll M$ while RG scale $\mu$ is not taken into account for a tree level analysis. Note  that $\Phi$ couples to SM through Higgs portal and the latter is a color singlet, this implies $c_{ggh}$ and $c_{gghh}$ are both zeros not only at tree level matching but at every order in the EFT. In other words for a scalar singlet model we can't have an effective Higgs-gluons coupling without integrating the the top quark out.\\
For the case of the color neutral scalar we do not get any contributions to $c_{ggh}$ and $c_{gghh}$. An explicit example where they arise will be discussed in section \ref{sec:coloredscalars}.
One can now do the exercise to understand where the SMEFT and HEFT couplings agree.
Using the results of Table \ref{tab::di-Higgs} and the relations among the parameters in Eqs.~\eqref{eq:mu4} one finds, expanding in large $m_2$ with respect to the other massive parameters and assuming in addition that $\theta$ is small,
\begin{align}
\frac{C_{H\Box}}{\Lambda^2}\approx & \frac{v_S \lambda_3}{v_H m_2^2}\theta -\frac{1}{2 v_H^2}\theta^2 +\frac{2 m_1^2 - 2 v_S^2 \lambda_2-5 v_H^2 \lambda_3^2}{2 v_H^2 m_2^2} \theta^2\,, \label{eq:CHBoxexp}\\
\frac{C_{H}}{\Lambda^2} \approx& \frac{\lambda_3}{2 v_H^2} \theta^2+\frac{v_H^2\lambda_3^2-m_1^2\lambda_3}{m_2^2 v_H^2}\theta^2\,. \label{eq:CHexp}
\end{align}
 Expanding in small $\theta$ and omitting in Eq.~\eqref{eq:CHBoxexp} and \eqref{eq:CHexp} terms of $\mathcal{O}(1/m_2^4)$ and $\mathcal{O}(\theta^2/m_2^2)$  we find the results in Table \ref{tab::matchingHEFTSMEFT}. Neglecting the terms of higher orders in $1/m_2^2$ indeed SMEFT and HEFT agree under the assumption of small mixing angles and $v_S\to 0$ (i.e.~assuming that $v_S v_H \ll m_2^2$). 
 \par
 HEFT matching is not unique \cite{Dawson:2023oce}. For instance one can adopt a different parameterisation of the singlet Lagrangian introducing a tadpole term but no $v_S$. In this case the matching to SMEFT does not contain the extra terms $\propto v_S$ and there is full agreement at $\mathcal{O}(\theta^2)$ \cite{Dawson:2020oco}. The parameters though get re-interpreted. The recent Ref.~\cite{Ge:2026qfa} points out that choosing a parameter set consisting of masses, mixing angle and vacuum expectation values, as we use here, leads to maximally precise matching results. Finally, we notice that in Refs.~\cite{Dawson:2023oce, Dittmaier:2021fls} which consider matching of the singlet model to HEFT further hierarchies among the UV parameters of the singlet model are imposed, which we do not need to impose here, nor are they necessary from the point of view of our power counting. As has been shown in Ref.~\cite{Buchalla:2016bse}, when assuming parametric hierarchies among the model parameters, the matching procedure directly determines whether the resulting low-energy description follows SMEFT or HEFT power counting. In particular, we notice that in the singlet model there is still parameter space allowed in which the linearisation of the mixing angle relation is not necessarily accurate. We are illustrating this, by showing in the following a comparison of couplings in HEFT and SMEFT relevant for Higgs pair production.
\begin{table}[!t]
\renewcommand*{\arraystretch}{2.5}
    \centering
\begin{tabular}{|c|c|c|}
   \hline
              &    HEFT       &   SMEFT                 \\
   \hline
   $c_{hVV}$  & $1-\frac{1}{2}\theta^2$   &  $1-\frac{1}{2}\theta^2 + \frac{v_S v_H\lambda_3}{m_2^2}\theta $
   \\
   \hline 
  $c_{hhVV}$ & $1-2 \theta^2$  &    $1- 2 \theta^2 + \frac{4 v_S v_H\lambda_3}{m_2^2}\theta $\\
   \hline
   $c_{hhh}$ & $1-\frac{3}{2}\theta^2 +\frac{\lambda_3 v_H^2}{m_1^2}\theta^2$  & $1-\frac{3}{2} \theta^2 +\frac{\lambda_3 v_H^2}{m_1^2}\theta^2+ \frac{3 v_H v_S \lambda_3}{m_2^2} \theta $  \\
   \hline
   $c_{t} $ & $1-\frac{1}{2}\theta^2 $  &  $1-\frac{1}{2}\theta^2 + \frac{v_S v_H\lambda_3}{m_2^2}\theta$  \\
   \hline
   $c_{2t}$ & $-\frac{1}{2}\theta^2$ &    $-\frac{1}{2}\theta^2 + \frac{v_S v_H\lambda_3}{m_2^2}\theta $ \\
    \hline
\end{tabular}
\caption{Relevant couplings in di-Higgs production up to $\mathcal{O}(\theta^3, \frac{\theta^2}{m_2^2}, \frac{1}{m_{h_2}^4}) $. }
\label{tab::matchingHEFTSMEFT}
\end{table}
\begin{figure}[!b]
\includegraphics[scale=0.51]{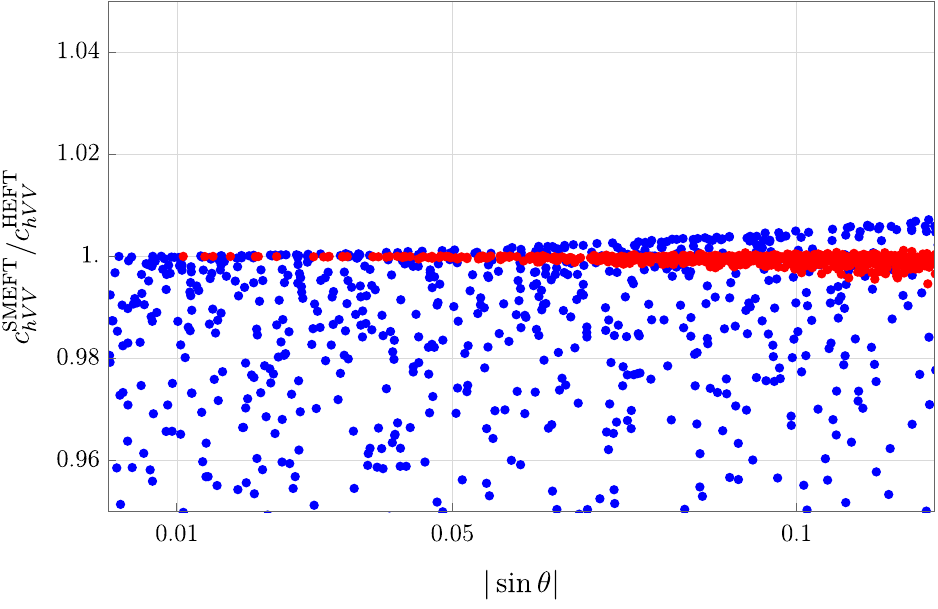}
\includegraphics[scale=0.51]{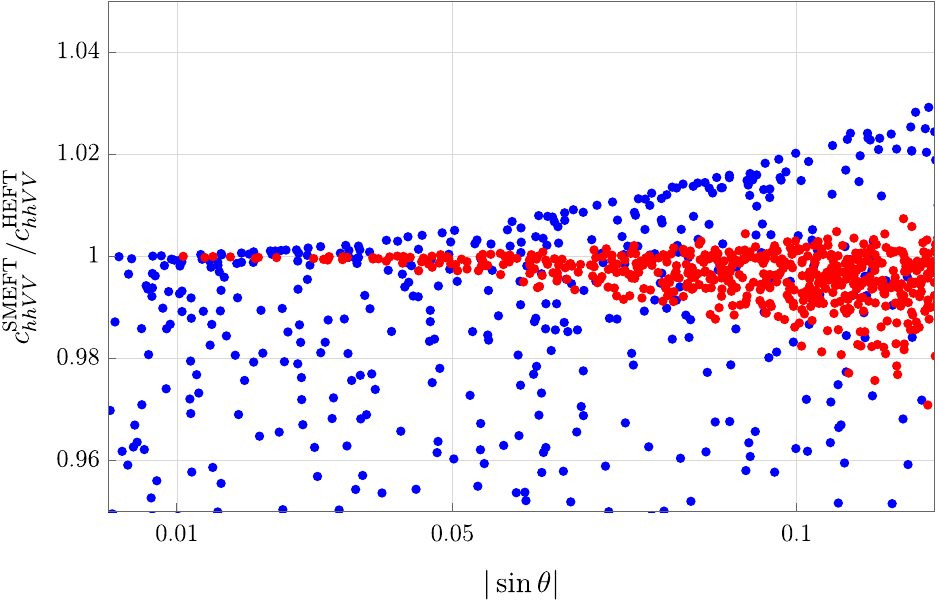}
\caption{Ratio of the SMEFT and HEFT predictions for the coupling modifiers $c_{hVV}$ and $c_{hhVV}$ for a scan over the parameter space, using $\lambda_2\in [0,4\pi]$, $\lambda_3\in [-4\pi, 4 \pi]$, $|\sin \theta| < 0.15$, $v_S\in [0.01, 5 v_H]$ and $M\in [2,3]\text{ TeV}$ (blue points). For the red points instead the range of $v_S$ was constrained to lie within $v_S\in [0.01, 0.1v_H]$. 
%\RG{in the figure we use capital $C$ while in the text everywhere $c$ }%}
\label{fig:cVVhandcVVh}}
\end{figure}
\par
The coupling modifiers $c_{hVV}$ and $c_{hhVV}$ are shown in Fig.~\ref{fig:cVVhandcVVh}, by scanning over the parameter space of the model imposing the conditions discussed in subsection~\ref{subsec:singlet_parameterscan}.  While the SMEFT and HEFT couplings agree well for small $v_S$ the disagreement becomes larger for large $v_S$. The disagreement for small $v_S$ points between SMEFT and HEFT is more pronounced for $c_{hhVV}$.
This can be understood from the fact that, at $\mathcal{O}(\theta^3)$, terms proportional to
$(v_H/v_S)\,\theta^3$ appear: for sufficiently small $v_S$, the enhancement by $v_H/v_S$ can
compensate the additional \(\theta\) suppression, yielding a non-negligible contribution.
While we do not show the plots for the other couplings, we note that we can see similar behaviour for $c_t$ and $c_{hhh}$ while the fact that we plot ratios, lead to a larger effect for $c_{2t}$ where no SM coupling exists.
We also comment on the fact that such a comparison on the level of couplings is not very physical. We could only do it as we were imposing the field redefinition in Eq.~\eqref{fieldref}, otherwise in SMEFT at the considered order also new Lorentz structures would have contributed. 

%Nevertheless If we keep a finite top quark mass and integrate it out we would  restrict the EFT range of validity to energies below top mass.\\ 
%Alternatively we can change our NP model, for instance a colored scalar, then we get non vanishing $c_{ggh}$ $c_{gghh}$.
%please correct the latter statement, It's something I was thinking about, so It may be wrong%
\subsubsection{Results for gluon fusion into a Higgs pair}
We compute inclusive cross section for Higgs pair production in gluon fusion using \texttt{hpair} \cite{hpair, Dawson:1998py, Grober:2015cwa} at LO in QCD. QCD corrections for gluon fusion Higgs pair production are large and their size depend on the EFT parameter point \cite{Buchalla:2018yce, Heinrich:2022idm}. But there is no public available code that allows to compute the NLO QCD corrections in full top mass dependence in the UV model. Adopting an infinite top mass limit the QCD corrections can be taken care of by a $K$ factor that in this limit is basically independent on the specifics of the EFT parameter point \cite{ Grober:2017gut, Grober:2015cwa}. Such a $K$ factor cancels in the ratios of cross sections we show in our figures.
What regards the UV model, we also need the width of the heavy Higgs boson. We compute it using the Higgs width for a Higgs boson with mass $m_2$ using \texttt{hdecay} \cite{Djouadi:1997yw, Djouadi:2018xqq} and add the partial width of the heavy Higgs boson decay to two light Higgs bosons given by
\begin{equation}
\Sigma_{h_2\to h_1 h_1}=\frac{1}{16\pi m_2}d_2^2\sqrt{1-4\frac{m_1^2}{m_2^2}}\,.
\end{equation}

We show our results as a function of $f$ which measures the amount of symmetry breaking to the mass of the new scalar.
Following \cite{Banta:2021dek} we define a $f$ parameter such that $0<f<1$ :
\[f = \frac{\lambda_3}{\lambda_3 + \frac{2\mu_{2}^2}{v_{H}^2}}=  \frac{\lambda_3}{\lambda_3 + \lambda_{\text{ex}}} \,. \] 
Moreover $f$ is closely related to the expansion parameter appearing when one solves the equation of motion of the scalar singlet to obtain the effective operator expansion $r_{\text{exp}}= \frac{\lambda_3 v_{H}^2}{\mu_{2}^{2}}$.
In fact these quantities may be related by:\[f = \frac{\lambda_3}{\lambda_3 + \frac{\mu_{2}^2}{2 v_{H}^2}}=  \frac{2 \lambda_3 v_{H}^2}{2 \lambda_3 v_{H}^2 +  \mu_{2}^2} = \frac{2r_{\text{exp}}}{1+ 2r_{\text{exp}}} \,. \] 
When $f\rightarrow 1 (r_{\text{exp}}\rightarrow \infty)$ the singlet acquires all its mass from electroweak symmetry breaking.
\par

\begin{figure}[!t]
    \centering
    \begin{minipage}[b]{0.49\textwidth}
        \includegraphics[scale=0.53]{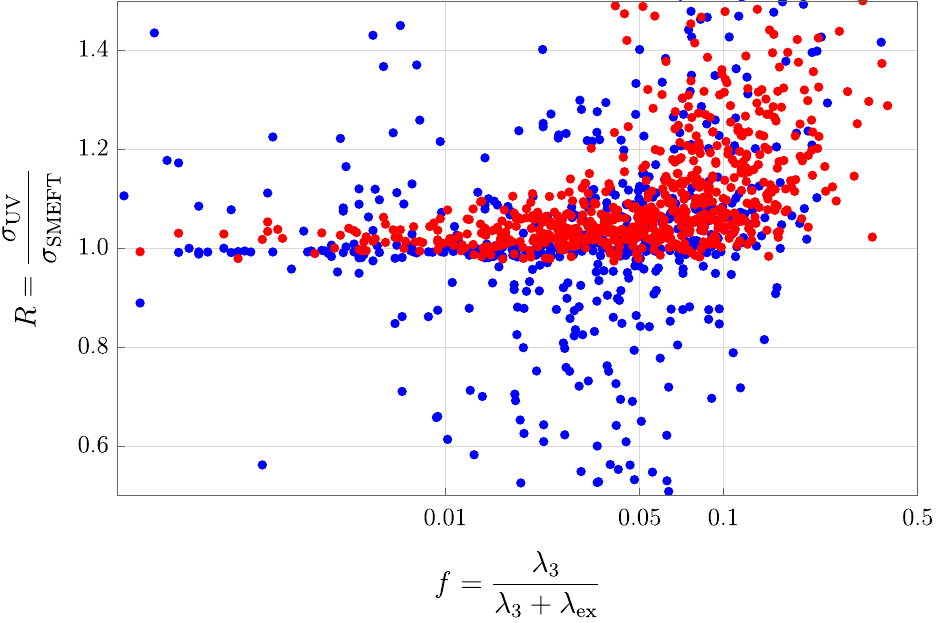}
        \label{fig:ggF_SMEFT_UV}
    \end{minipage}
    \hfill
    \begin{minipage}[b]{0.49\textwidth}
        \includegraphics[scale=0.53]{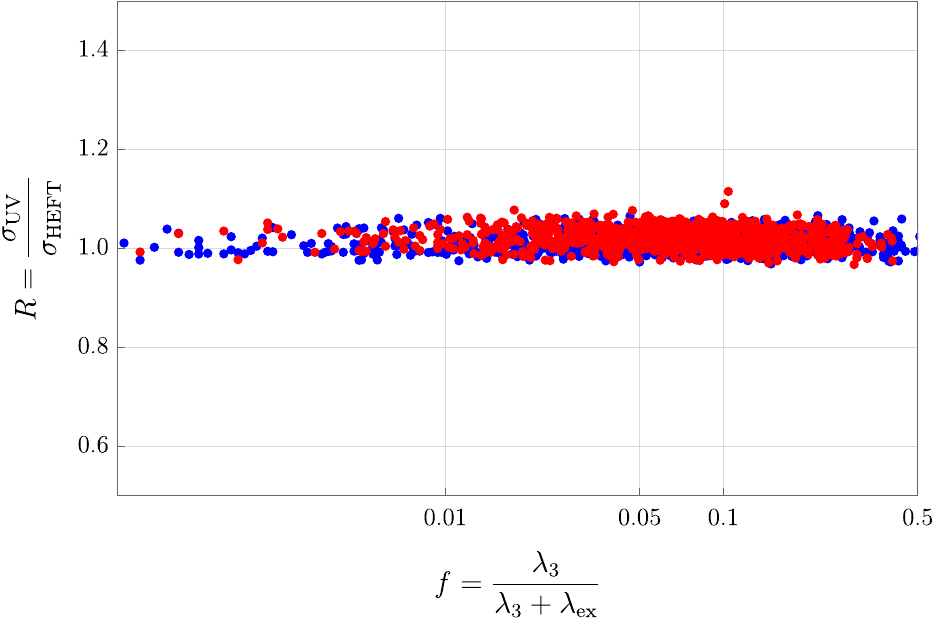}
        \label{fig:ggF_HEFT_UV}
    \end{minipage}
    \caption{Cross section for Higgs production in gluon fusion using the scalar singlet model. On the right (left) panel we display ratio of $\sigma_{\text{UV}}$ over the same quantity computed in HEFT (SMEFT) framework. Red (blue) points refer to scan 1 (scan 2) which allows for small values only (also order one values) of $v_S/v_H$.}
    \label{fig::xs for ggF in the scalar singlet}
\end{figure}

The results are collected in Fig.~\ref{fig::xs for ggF in the scalar singlet}. The red and blue points originate from different parameter scans, imposing the bounds from subsection~\ref{subsec:singlet_parameterscan} and 
scanning as follows:
\begin{equation}
 \lambda_{2}\in [0,4\pi],\quad  \lambda_{3}\in [-4\pi, 4\pi], \quad |\sin \theta|\leq 0.15, \quad m_2\in [800, 3000] \text{ GeV}, \label{eq:scan}
\end{equation}
and
\begin{itemize}
    \item scan 1:  $v_{S} \in[0.01, 0.1\,v_{H}]$,
    \item scan 2:  $v_{S} \in[0.01, 5\,v_{H}]$.
\end{itemize}
While the values for the $\lambda$'s , the mixing angle and $v_{S}$ are uniformly distributed over the range given in Eq.~\eqref{eq:scan}, the values of $m_2$  were chosen between normal distributions with mean $1800, 2200, 2600$ GeV and standard deviation of $400$ GeV. 
Since we truncate the cross section at $\mathcal{O}(1/\Lambda^2)$ it can happen that $\sigma_{\text{SMEFT}}<0$. We interpret this as a breakdown of the SMEFT approximation \cite{Heinrich:2022gzl}.
In particular, we notice that this can happen in presence of high values of $\lambda_3$ i.e. high $f$ values, implying that we are outside the convergence radius of SMEFT. The points with negative cross section are not shown in the plot as the $y$ axis shows only positive values of $R=\sigma_{UV}/\sigma_{SMEFT}$.
Figure~\ref{fig::xs for ggF in the scalar singlet} clearly shows that HEFT is the better EFT description for the considered model and parameter ranges. SMEFT can provide an accurate description if $v_S$ and $f$ are small.

\subsubsection{Results for vector boson fusion into a Higgs pair}
Vector boson fusion is the second biggest contribution for Higgs pair production at LHC. New physics effects dominantly affect the hard subprocess $V V \rightarrow h h$ where $V = W^{\pm},Z$, while the forward tagging jets mostly provide a kinematic handle for event selection. We will hence concentrate on $V V \rightarrow h h$ and compare its unpolarized amplitude squared computed in the scalar singlet model with the same quantity computed in the EFT frameworks SMEFT and HEFT.\footnote{For a similar study in the context of the triplet extension of the SM, see Ref.~\cite{Song:2025kjp}.} 
We computed the amplitudes at different centre of mass (c.o.m.)  energies $\sqrt{s}$ at fixed value of the scattering angle. Since no significant qualitative difference has shown up at different values of $\sqrt{s}$ we refrain to compute the full cross sections.
We also note that the QCD corrected cross sections are basically insensitive to the specifics of the EFT parameter point \cite{Braun:2025hvr, Jager:2025isz}.
\par
\begin{figure}[!t]
    \flushleft
    \begin{minipage}[b]{0.49\textwidth}
        \includegraphics[scale=0.53]{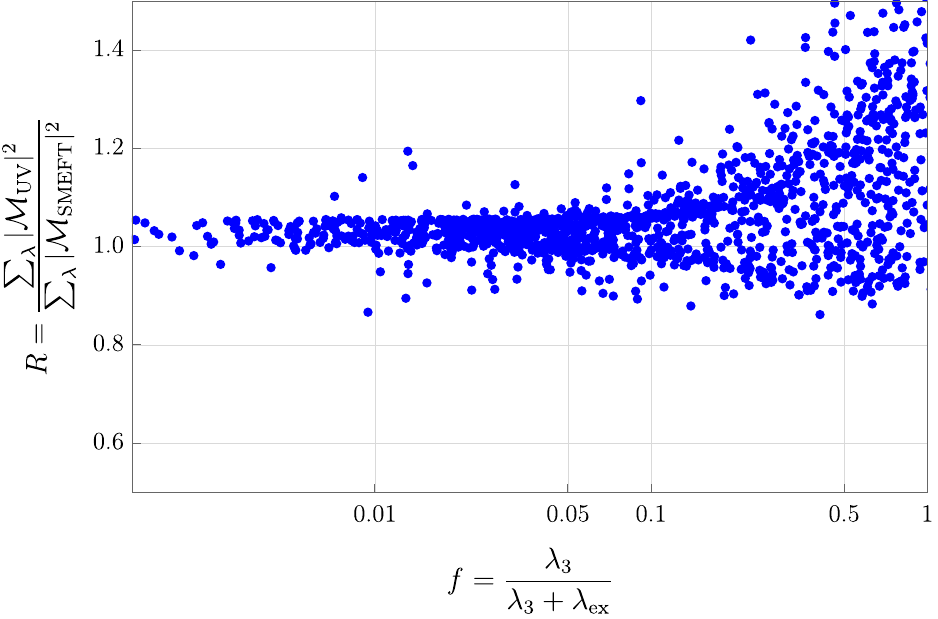}
        \label{fig:image1}
    \end{minipage}
    \hfill
    \begin{minipage}[b]{0.49\textwidth}
        \includegraphics[scale=0.53]{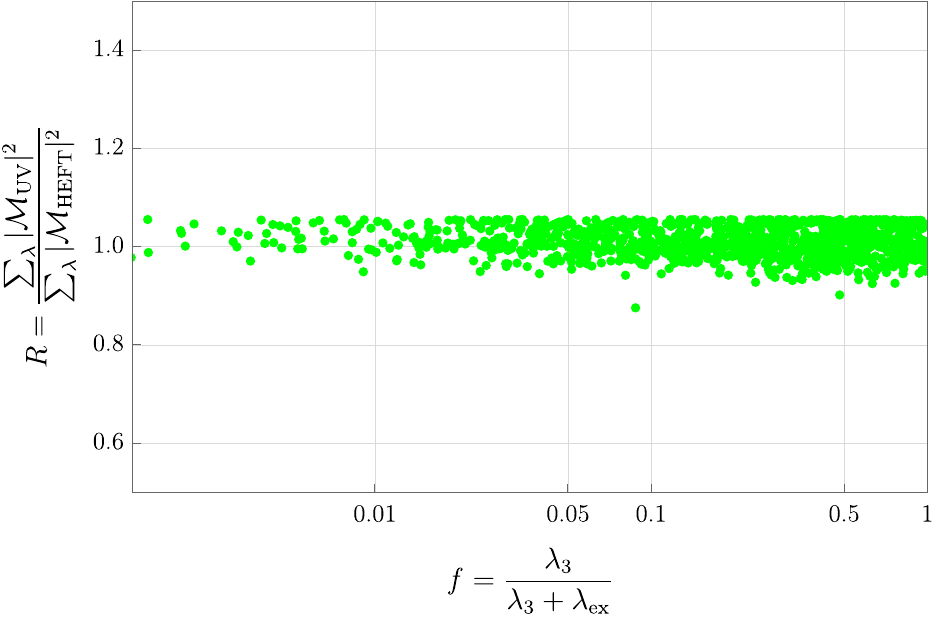}
        \label{fig:image2}
    \end{minipage}
    \caption{Unpolarized amplitudes squared for Higgs pair production in vector boson fusion for the scalar singlet model. On the right (left) panel we display the ratio of the unpolarized amplitude squared in the UV model over the same quantity computed in HEFT (SMEFT) framework .}
    \label{fig::VBF amplitudes squared in the scalar singlet}
\end{figure}

Our results are summarized in Fig.~\ref{fig::VBF amplitudes squared in the scalar singlet}. 
The points displayed in the figure result from a scan in the parameter space similar to the one in Eq.~\eqref{eq:scan}
but using 
\begin{equation}
v_S \in [0.2v_{H},  2 v_H]\,.
\end{equation}
Scans which allow higher values of $v_S$ follow a very similar behaviour as the one shown. In Fig.~~\ref{fig::VBF amplitudes squared in the scalar singlet} we have fixed the scattering angle to its maximal value $\alpha_{\text{max}}= \frac{\pi}{2}$ and $\sqrt{s}=1 $ TeV. 
The SMEFT amplitudes $\mathcal{M}_{\text{SMEFT}} = \mathcal{M}_{\text{SM}} + \frac{c}{\Lambda^2} \mathcal{M}^{(6)}$ were computed under linearised SMEFT approximation $|\mathcal{M}_{\text{SMEFT}}|^2 = |\mathcal{M}_{\text{SM}}|^2 + 2 \frac{c}{\Lambda^2} \text{Re}(\mathcal{M}_{\text{SM}}\mathcal{M}^{(6)}) +\mathcal{O}\left( 1/\Lambda^4\right)$ i.e. neglecting dimension-8 contributions stemming from the matrix element squared. We checked though explicitly that basically no difference emerges when including those terms of $\mathcal{O}\left( 1/\Lambda^4\right)$ stemming from $\text{Re}(\mathcal{M}_{\text{SM}}\mathcal{M}^{(6)})$.\par
From Fig.~\ref{fig::VBF amplitudes squared in the scalar singlet} we interfere that while for $f<0.1$ SMEFT and HEFT do nearly equally well, for larger values of $f$ HEFT seems to describe the full UV model better.
While Fig.~\ref{fig::VBF amplitudes squared in the scalar singlet} is for fixed $\sqrt{s}$ and scattering angle, one can observe that for higher (lower) values of $\sqrt{s}$ SMEFT describes the UV model well up to higher (lower) values of $f$ with respect to $f=0.1$, HEFT does not show significant deviation.

\subsection{Two Higgs Doublet Model}
The two Higgs doublet model (2HDM) \cite{Branco:2011iw} is constructed by adding a second $SU(2)$ doublet $\Phi_2$ to the spectrum of the SM. In the following we will slightly change notation and nominate both the Higgs doublets by $\Phi_i$ with $i=1,2$. 
We assume CP-conservation and impose a $\mathbb{Z}_2$ symmetry on the potential that acts as $\Phi_1\to \Phi_1$ and $\Phi_2\to -\Phi_2$. This symmetry assignment avoids flavour-changing neutral currents and is softly-broken by an off-diagonal mass term $m_{12}$.
The spectrum of the 2HDM contains two CP-even neutral Higgs bosons $h$ and $H$, a CP-odd Higgs boson $A$ and two charged Higgs bosons $H^{\pm}$.
The potential can then be written as
\begin{equation}
\begin{split}
V(\Phi_1, \Phi_2)=&m_{11}^2 (\Phi_1^{\dagger} \Phi_1)+m_{22}^2 (\Phi_2^{\dagger} \Phi_2)-m_{12}^2\left[\Phi_1^{\dagger} \Phi_2+\text{h.c.}\right]+\frac{\lambda_1}{2} (\Phi_1^{\dagger} \Phi_1)^2+\frac{\lambda_2}{2} (\Phi_2^{\dagger} \Phi_2)^2\\ + & \lambda_3 (\Phi_1^{\dagger} \Phi_1) (\Phi_2^{\dagger} \Phi_2)
+\lambda_4 (\Phi_1^{\dagger}\Phi_2) (\Phi_2^{\dagger}\Phi_1)+\frac{\lambda_5}{2}\left[(\Phi_1^{\dagger}\Phi_2)^2+\text{h.c.}\right]\,. \label{eq:V2HDM}
\end{split}
\end{equation}
Each of the doublets can acquire a vacuum expectation value $\langle \Phi_1 \rangle=v_1/\sqrt{2}$ and $\langle \Phi_2 \rangle=v_2/\sqrt{2}$ with $v^2=v_1^2+v_2^2=(246 \text{ GeV})^2$. The angle $\beta$ with $\tan\beta=v_2/v_1$ diagonalises the charged  pseudoscalar Higgs sector. It is useful to rotate the original doublets into the \textit{Higgs basis} 
\begin{equation}
    \begin{pmatrix}
        H_1 \\
        H_2
    \end{pmatrix} 
    =
    \begin{pmatrix}
        c_\beta & s_\beta \\
        -s_\beta & c_\beta
    \end{pmatrix} 
    \begin{pmatrix}
        \Phi_1 \\
        \Phi_2
    \end{pmatrix},
\end{equation}\label{eq:2hdm_rotation} 
with $c_{\beta}=\cos\beta$ and $s_{\beta}=\sin\beta$,
in which only one of the doublets acquires a vacuum expectation value, namely $\langle H_1 \rangle=v/\sqrt{2}$ and $\langle H_2\rangle=0$.
The potential in the Higgs basis reads
\begin{equation}
\begin{split}
V(H_1, H_2)=&M_{11}^2 (H_1^{\dagger} H_1)+M_{22}^2 (H_2^{\dagger} H_2)-M_{12}^2\left[H_1^{\dagger} H_2+\text{h.c.}\right] \\+ &\frac{\Lambda_1}{2} (H_1^{\dagger} H_1)^2+\frac{\Lambda_2}{2} (H_2^{\dagger} H_2)^2 +  \Lambda_3 (H_1^{\dagger} H_1) (H_2^{\dagger} H_2)
+\Lambda_4 (H_1^{\dagger}H_2) (H_2^{\dagger}H_1) \\ +&\left[\frac{\Lambda_5}{2}(H_1^{\dagger}H_2)^2+\Lambda_6 (H_1^{\dagger}H_1)(H_1^{\dagger}H_2)+\Lambda_7 (H_1^{\dagger}H_2)(H_2^{\dagger}H_2)+\text{h.c.}\right]\,. \label{eq:HiggsbasisV2HDM}
\end{split}
\end{equation}

The relations between the parameters of the Higgs basis potential and the potential in Eq.~\eqref{eq:V2HDM} can be found in full generality in Appendix A of Ref.~\cite{Davidson:2005cw}.
In order to obtain the physical Higgs bosons, a rotation
\begin{equation}
    \begin{pmatrix}
        H \\
        h
    \end{pmatrix} 
    =
    \begin{pmatrix}
        c_{\beta-\alpha} & s_{\beta-\alpha} \\
        -s_{\beta-\alpha} & c_{\beta-\alpha}
    \end{pmatrix} 
    \begin{pmatrix}
        H_1^0 \\
        H_2^0
    \end{pmatrix},\label{eq:2hdm_rotation2}
\end{equation}
where $H_i^0$ denotes the neutral CP-even component of $H_i$, is performed.
Finally, we note that all the physical heavy Higgs boson masses $m_{H^{\pm}}$, $m_A$ and $m_{H}$ obtain a component of an explicit mass term and a mass component from electroweak symmetry breaking.
For instance we can write
\begin{equation}
m_{H}^2=M^2+ \tilde{\Lambda} v^2\label{eq:mHTHDM}
%=M^2+\Delta m_H^2 \label{eq:mHTHDM}
\end{equation}
where  $\tilde{\Lambda}$ is a combination of the $\lambda_i$ (see e.g.~Ref.~\cite{Degrassi:2023eii}) and
\begin{equation}
M^2=\frac{m_{12}^2}{\sin\beta \cos\beta}\,.
\end{equation}
We use as input parameters
\begin{equation}
m_h, \quad m_{H}, \quad m_A, \quad m_{H^{\pm}}, \quad t_{\beta}=\tan\beta,\quad c_{\beta-\alpha},\quad M^2\,.
\end{equation}
%EFT for 2HDM.  \cite{Dermisek:2024ohe}
\subsubsection{Matching to HEFT and SMEFT}
We match the 2HDM to HEFT and SMEFT.
The matching to SMEFT is well known and has been computed up to dimension 8 in \cite{Dawson:2022cmu} and at one-loop order in \cite{DasBakshi:2024krs}\footnote{See also \cite{Banta:2023prj} for the choice of an especially convenient choice of basis.}. For the couplings relevant for Higgs pair production in gluon fusion, one finds, matching at tree level \cite{DasBakshi:2024krs}, 
\begin{align}
\frac{C_{tH}}{\Lambda^2}=&\frac{y_u \Lambda_6}{M^2 \tan\beta}=-\frac{\sqrt{2}}{v^3} \frac{c_{\beta-\alpha}}{t_{\beta}} m_t-\frac{1}{\sqrt{2}v^3}\left(c_{\beta-\alpha}^2 +\frac{c_{\beta-\alpha}(4 \Delta m_H^2-6 m_h^2)}{t_{\beta}\Lambda^2}\right) m_t \,, \label{eq:THDMCtH}\\
\frac{C_{H}}{\Lambda^2}= &=\frac{\Lambda^2}{v^4} c^2_{\beta-\alpha}+\frac{4}{v^4}c^2_{\beta-\alpha}(\Delta m_H^2-m_h^2)\,.
\end{align}
$\Lambda$ takes the role of the heavy masses that are assumed to be degenerate and 
\begin{equation}
\Lambda^2=M^2-\frac{1}{2}\left(s_{\beta-\alpha}^2 m_h^2+c_{\beta-\alpha}^2 m_H^2 \right)-c_{\beta-\alpha} s_{\beta-\alpha}\tan (2\beta) (m_h^2-m_H^2)\,,
\end{equation}
whereas $\Delta m_H^2=m_H^2-\Lambda^2$. Instead for HEFT we find
\begin{align}
&c_t=s_{\beta-\alpha}+\frac{c_{\beta-\alpha}}{t_{\beta}}\,,\\
%c_V=&s_{\beta-\alpha}\,,\\
&c_{hhh}=\frac{2v}{m_1^2}d_1^{\text{2HDM}}\,, \label{eq:chhh2HDM}\\
&c_{2t}=-\frac{(c_{\beta-\alpha}-\frac{s_{\beta-\alpha}}{t_{\beta}}) v d_2^{\text{2HDM}}}{m_H^2}\,, \label{eq:c2t2HDM}
\end{align}
with $d_i^{\text{2HDM}}$ given in Appendix~\ref{app:couplings}. In analogy to our considerations for the singlet model, we want to understand in which limit the couplings correspond to each other in SMEFT and HEFT. We expand in small mixing angle, noting that the SMEFT limit of large Lagrangian mass corresponds to small mixing between light an heavy Higgs bosons, $c_{\beta -\alpha}\ll 1$. The results of this expansion can be found in Table~\ref{tab:THDMsmallmixing}.
\begin{table}[htbp]
\begin{center}
\renewcommand*{\arraystretch}{2.5}
    \centering
\begin{tabular}{|c|c|c|}
   \hline
              &    HEFT       &   SMEFT                 \\
   \hline
   $c_{hhh}$ & $1-2\frac{M^2}{m_h^2}c^2_{\beta-\alpha}$  &  $1-2\frac{4m_H^2-3 M^2}{m_h^2}c^2_{\beta-\alpha}$  \\
   \hline
   $c_{t} $ & $1+\frac{c_{\beta-\alpha}}{t_{\beta}}$ & $1+\frac{c_{\beta-\alpha}}{t_{\beta}}+2 \frac{m_H^2-M^2}{t_{\beta}M^2}$   \\
   \hline
   $c_{2t}$ & $4\frac{M^2}{m_{H}^2}\frac{c_{\beta-\alpha}}{t_{\beta}} -\frac{c_{\beta-\alpha}}{t_{\beta}}$  &   $3\frac{c_{\beta-\alpha}}{t_{\beta}} +6 \frac{m_H^2-M^2}{t_{\beta} M^2} $ \\
    \hline
\end{tabular}
\caption{Relevant couplings in di-Higgs production up to $\mathcal{O}(c_{\beta-\alpha})$ for the 2HDM. $c_{hVV}$ becomes equal to one at the considered order. The trilinear Higgs self-coupling has been expanded an order higher in $c_{\beta-\alpha}$, see text. \label{tab:THDMsmallmixing}}
\end{center}
\end{table}

We note that even in the limit of small $c_{\beta-\alpha}$ the functional forms of $c_{hhh},\,c_t,$ and $c_{2t}$ in HEFT and SMEFT is different. In the limit, $v\to 0$ and using Eq.~\eqref{eq:mHTHDM} we see though that the SMEFT and HEFT coupling agree, again clearly showing that SMEFT is an expansion in a large Lagrangian parameter, here $M^2$, while HEFT expands in large physical mass. 
In the following, we will compare HEFT and SMEFT predictions in gluon fusion, while we expect in vector boson fusion less interesting effects due to the suppression of the effects in Higgs vector-boson couplings, starting only at order $c_{\beta-\alpha}^2$.
%%%%%%%%%%%%%%%%%%%%%%%%%%%
\subsubsection{Theoretical and experimental constraints on the model \label{sec:THDMconstraints}}
In order to take into account constraints on the parameter space we follow the discussion of Ref.~\cite{Arco:2020ucn}.
%We can fulfill constraints from electroweak precision tests trivially by assuming a degeneracy in the Higgs masses which do not enter into our analysis, namely by setting $m_A=m_{H^{\pm}}$. 
\begin{description}
\item[Vacuum stability:] We require boundedness from below, leading to the criteria 
\begin{align}
\lambda_1 \geq 0\,,\\
\lambda_2 \geq 0\,, \\
\lambda_3 +\sqrt{\lambda_1 \lambda_2} \geq 0\,,\\
\lambda_3 + \lambda_4 -|\lambda_5|+\sqrt{\lambda_1 \lambda_2} \geq 0 \,,
\end{align}
implemented as a required condition when scanning over the parameter space.
\item[Perturbativity:] Requiring that the eigenvalues of the lowest partial wave matrix remain below $16\pi$ puts bounds on the values of the scalar potential $\lambda_i$. The conditions can be found in \cite{Arco:2020ucn} and we implement them as a required condition to keep a point in a scan over the parameter space. 
\item[Electroweak precision tests:] Assuming that the heavy Higgs boson masses are degenerate either $m_A \sim m_{H^{\pm}}$ or  $m_H\sim m_{H^{\pm}}$ avoids constraints from electroweak precision tests. Since we do not need to specify the masses $m_A$ or $m_{H^{\pm}}$ in our analysis we can trivially satisfy the constraints imposed by electroweak precision tests.
\item[Flavour physics:] Flavour physics, in particular the decays $B_s\to X_s \gamma$ and $B_s\to \mu^+ \mu^-$ can impose important bounds on the value of $\tan\beta$ for light and moderate values of $m_{H^{\pm}}$. Those limits depend on the type of 2HDM considered. Since we consider the EFT limit, assuming that the heavy Higgs boson masses are larger than 1 TeV, we can evade them both in the type I and type II 2HDM by imposing in our scan that $\tan\beta > 0.9 $. 
\item[Higgs coupling measurements:] In order to assure that the SM-like Higgs boson fulfills constraints from experimental Higgs coupling measurements, we use the results of the ATLAS SMEFT scan in Ref.~\cite{ATLAS:2024lyh}. This means that we match to the SMEFT at tree level and keep a point in parameter space if the $\chi^2< \chi^2_{min}+5.99$. The $\chi_{min}$ is determined within the physical allowed range for $\tan\beta$ (see below).
\end{description}
We generate points in parameter space that fulfill these conditions, scanning over the parameter space with
\begin{equation}
c_{\beta-\alpha}\in[-0.2, 0.2]\,, \quad \tan\beta \in [0.9, 50]\, ,\text{GeV}^2\,, \quad m_H\in[1000, 3000]\,\text{GeV}\,.
\end{equation}
We restrict the value of $c_{\beta-\alpha}$ to the shown range to improve the effectiveness of the scan taking into account already typical constraints stemming from Higgs coupling measurements. We generate $M^2$ in a Gaussian distribution around $m_H^2\cos^2\alpha/\sin^2\beta$ \cite{Arco:2020ucn} to obtain more efficiently points that fulfill the requirement of stability and perturbativity constraints.
Those depend also on the masses $m_A$ and $m_{H^{\pm}}$ that we in principle do not need to specify for our analysis. We address this by setting $m_A=m_{H^{\pm}}$ as required by electroweak precision tests and scan in steps the remaining parameter to check whether perturbativity and partial wave unitarity can be fulfilled for any value of $m_{H^{\pm}}^2\in [10^6, 10^9] \,\text{GeV}^2$.
 We consider two realisations of the 2HDM, the type I and the type II model. In our parameter scan, the only difference in the model arises from the Higgs coupling measurements. While in the type I model the Yukawa type operator $C_{fH}$ is the same for up-type, down-type and leptons (the only difference in Eq.~\eqref{eq:THDMCtH} is that the top mass is to be replaced by the respective fermion mass), in the type II model the Yukawa-type operator of down-type fermions and leptons is given by
 \begin{equation}
\frac{C_{fH}}{\Lambda^2}=\frac{\sqrt{2}}{v^3} c_{\beta-\alpha} t_{\beta} m_f-\frac{1}{\sqrt{2}v^3}\left(c_{\beta-\alpha}^2 -\frac{t_{\beta}c_{\beta-\alpha}(4 \Delta m_H^2-6 m_h^2)}{\Lambda^2}\right) m_f \,, \label{eq:THDMCbH}
 \end{equation}
 with $f=b,s,d,\tau,\mu,e$.
 As a consequence a much smaller parameter space survives the bounds on Higgs coupling measurements in the type II model with respect to the type I model \cite{ATLAS:2024lyh}. In particular one is constrained to smaller values of $c_{\beta-\alpha}$ and $t_{\beta}$.

%\item some comment on flavour, flavour basically excludes $\tan\beta > 1.9$
\subsubsection{Results for gluon fusion into a Higgs boson pair}
\begin{figure}
\centering
\includegraphics[scale=0.51]{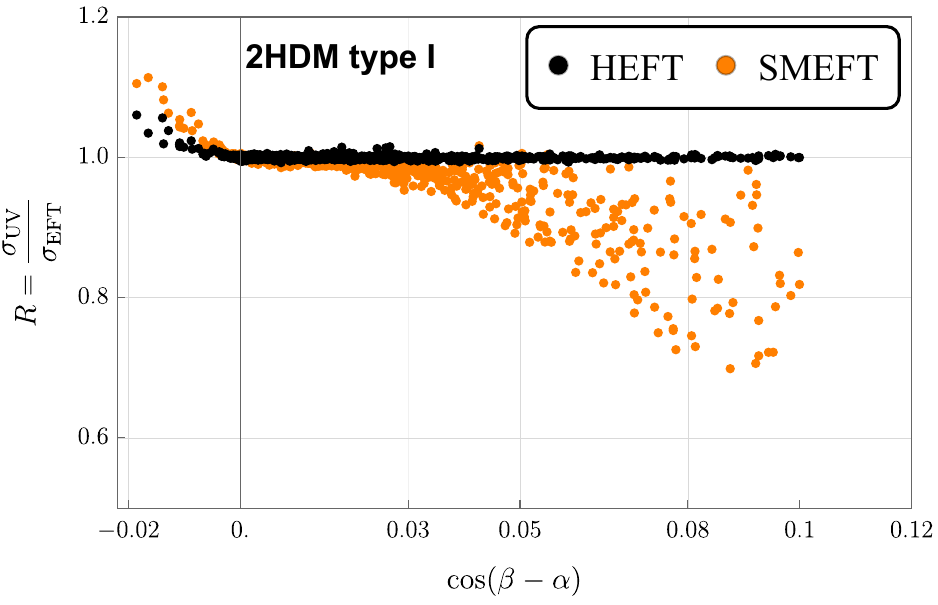}
\includegraphics[scale=0.51]{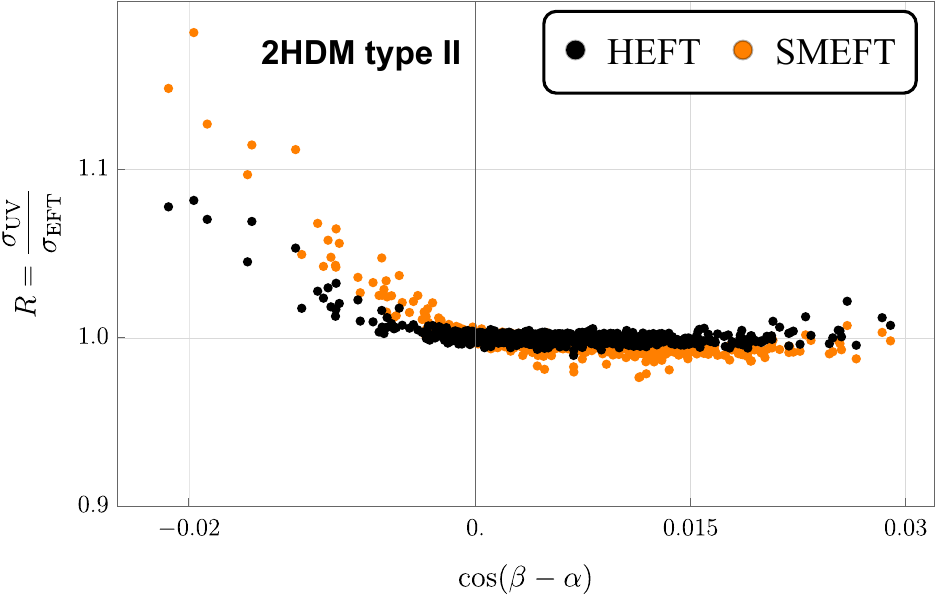}
\caption{Ratio of the full model cross section over the EFT cross section, for the 2HDM of type I (left) and type II (right). The black 
points show the case of matching to the HEFT and the orange points show the case of matching to the SMEFT. 
\label{fig:THDMcxn}}
\end{figure}

In Fig.~\ref{fig:THDMcxn} we present the result for the type I and type II model. The points are compatible with the constraints one the parameter space imposed in subsection~\ref{sec:THDMconstraints}. In order to compute the heavy Higgs width, needed for the UV model predictions, we have used \texttt{hdecay} \cite{Djouadi:1997yw, Djouadi:2018xqq}. 
In the cross section computation, we have only considered the top quark loops, which is motivated by the fact that in the SM the bottom loops are strongly suppressed \cite{Dolan:2012rv}. They could play a larger role in the type II model at large $t_{\beta}$. Since points with a large enhancements in the bottom couplings are though excluded by Higgs coupling measurements, this is a safe approximation. 

As can be inferred from Fig.~\ref{fig:THDMcxn}, the type II model allows for larger deviations from the alignment limit and as such we also find larger differences between the SMEFT and HEFT descriptions. The HEFT describes the UV model very well for positive values of $c_{\beta-\alpha}$, i.e. the ratio shown in the plot by the black points is very close to 1. In SMEFT, the ratio between EFT and UV result as shown by the orange points which deviate more and more from 1 with increasing value of $c_{\beta-\alpha}$. The effect is much smaller in the type II model as $c_{\beta-\alpha
}$ is much more constrained.
For our points with negative signs of $c_{\beta-\alpha}$ also the sign of the coupling of the heavy Higgs boson to top quarks is flipped, whereas the heavy to light Higgs self coupling remains positive hence changing the interference with the continuum in $gg\to hh$.
Having a delicate interference structure, amplifies the mismatch of truncated EFT descriptions for $c_{\beta-\alpha}<0$.

\begin{figure}
\centering
\includegraphics[scale=0.5]{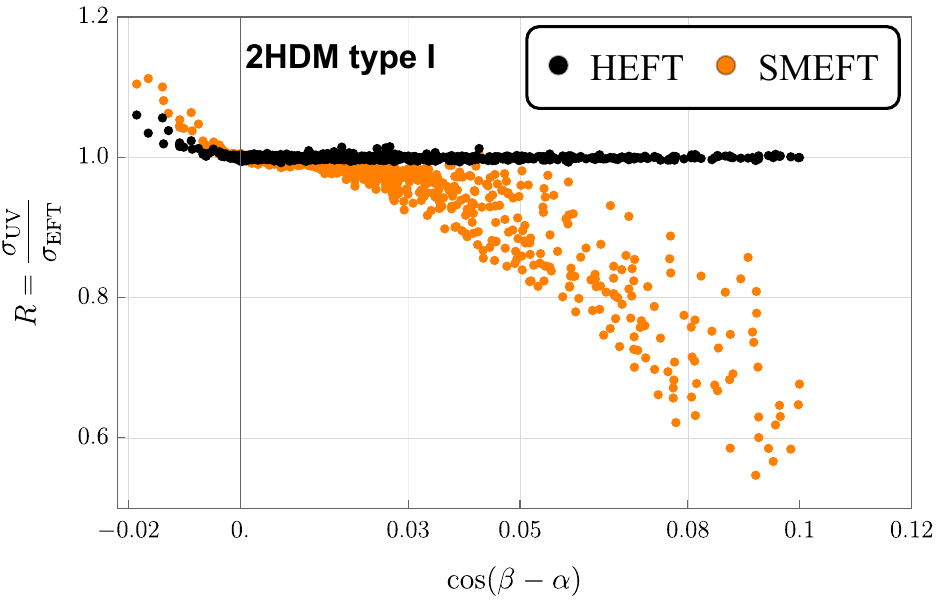} 
\caption{Ratio of the full model cross section over the EFT cross section, for the 2HDM of type I truncating the SMEFT at $\mathcal{O}(1/\Lambda^2)$ at the matrix element level, while keeping quadratic terms in the cross section.  %\RG{to be redone}%
\label{fig:THDMcxnquadratic}}
\end{figure}

For the SMEFT, we have considered only orders of to $1/\Lambda^2$ in the cross section. In EFT analysis, sometimes SMEFT is instead truncated at order $1/\Lambda^2$ at the level of the matrix element, allowing hence terms of order $1/\Lambda^4$ in the cross section. 
In Fig.~\ref{fig:THDMcxnquadratic} we show the same as the left plot of Fig.~\ref{fig:THDMcxn} for the type I model but truncated at the level of the matrix element. We see that indeed this choice worsens the agreement between the EFT and UV descriptions of the model.
\par
A comment on the choice of $c_{\beta-\alpha}$ as $x$ axis is in order. Similar to the singlet case, one could have defined a parameter $f$ that measures how much mass comes from electroweak symmetry breaking and how much from an explicit mass parameter in the Lagrangian. In that respect, for instance an $f=(m_H^2-M^2)/M^2$ could have been defined. We note however given our choice of $m_H$ to be above 1 TeV to remain in an EFT regime and imposing vacuum stability and perturbativity automatically selects only small values of $f$ making $c_{\beta-\alpha}$ the decisive parameter.\footnote{This is also in line with the findings of \cite{Kilic:2026ogm} which shows that Loryons with values of $f>0.5$ have masses up to 700 GeV. We note though that our model is a bit different from the one in the study of \cite{Kilic:2026ogm} as they have no soft breaking of the $\mathbb{Z}_2$ symmetry.}
\par
Finally, we note that we have performed the matching only at tree level. It is well known that loop corrections to the trilinear Higgs self-couplings can be largely enhanced with respect to the tree-level in the 2HDM even in the alignment limit $c_{\beta-\alpha}\to 0$ in the presence of large scalar potential couplings, see e.g.~\cite{Kanemura:2002vm, Kanemura:2004mg, Braathen:2019pxr, Braathen:2019zoh, Bahl:2023eau, Degrassi:2025pqt}. In this respect, it might be interesting to see how well HEFT describes the full model including also higher orders both in the matching and the full model. Since even the two-loop corrections are large, this is clearly beyond the scope of this work.

\subsection{Colored scalars \label{sec:coloredscalars}}
In this section we study a colored scalar $\omega_1$ whose charges under SM group are $\left(3,1\right)_{-1/3}$, where the numbers indicate the $(SU(3), SU(2)_L)_Y$ representations or charges. The SM Lagrangian is extended by
\begin{align}
\mathcal{L}\supset D_{\mu}\omega_1^\dagger D^{\mu}\omega_1- M_{\text{ex}}^2\omega_1^\dagger\omega_1-\frac{c_{\lambda\phi}}{2}\omega_1^{\dagger}\omega_1 H^{\dagger}H.
\label{eq:lagw1}
\end{align}
In Eq.~\eqref{eq:lagw1}, the first term contains the covariant derivative which couples $\omega_1$ to the gauge sector, the second piece is the explicit mass term and the last one represents a Higgs-portal interaction which provides another contribution to $\omega_1$ mass in the broken phase. The Feynman rules generated by Eq.~\eqref{eq:lagw1} relevant for our computation are presented in Appendix \ref{app:Feynman Rules}.

\subsubsection{Effective field theory for the colored scalar}
Here we compute the contributions to the effective operators affecting gluon fusion Higgs production. 
%leads 
%to a finite shift to the Higgs mass $\abs{H}^2$ as well as to the Higgs self couplings $\abs{H}^4$ from the loop corrections.
%\begin{equation}
%    \mu_{1}^2 \rightarrow \mu_{1}^2 + \frac{3}{32\pi^2} c_{\lambda\phi} M_{\text{ex}}^2 \,, \quad \lambda_{H}\rightarrow \lambda_{H}- \frac{45 c_{\lambda\phi}^2 +g_{1}^4}{180 (4\pi)^2 M_{\text{ex}}^2} \mu_{1}^2 \,.
%\end{equation}
What regards the matching of the color scalar model, we stay strictly at the one-loop order, so keep the one-loop matching contributions to $\mathcal{O}_{HG}$ as the operator enters at tree level, but neglect the effect of one-loop matching to $C_{H}$ as this would be effectively be a two-loop contribution to the gluon fusion process.
%As far as gluon fusion is concerned we generate a $\mathcal{O}_{HG}$ which gives a contribution to the lowest order SMEFT analysis. \RG{Question: does the model at one-loop also get contributions to $C_{H}$ and $C_{tH}$? If yes we should leave away the discussion of the $\mu_1$ and $\lambda_H$ and instead write that we don't consider those employing a loop-counting argument (they enter as one-loop matched contributions into one-loop diagrams)}
%\LT{I've checked the Matchete file: $C_H$ is generated, but apparently $C_{tH}$ vanishes.}\\
The beyond-the SM scattering amplitude in SMEFT up to dimension 6, cross-checked with \texttt{SMEFTsim} \cite{Brivio:2020onw}, is
\begin{align}
\mathcal{M}_{SMEFT}(gg \rightarrow hh)&=\frac{24 \, C_{HG} \, \lambda_H \, {v}_{H}^{2} \, 
\delta^{\text{ab}} \, 
\varepsilon^{\mu}(p_{1}) \, 
\varepsilon^{\nu}(p_{2}) 
\left( p_{2}^{\mu} p_{1}^{\nu} 
- g^{\mu\nu} (p_{1}\cdot p_{2}) \right)}
{\Lambda^{2} \left( s - m_{H}^{2} \right)} \nonumber \\
&\quad + \frac{4 \, C_{HG} \, 
\delta^{\text{ab}} \, 
\varepsilon^{\mu}(p_{1}) \, 
\varepsilon^{\nu}(p_{2}) 
\left( p_{2}^{\mu} p_{1}^{\nu} 
- g^{\mu\nu} (p_{1}\cdot p_{2}) \right)}
{\Lambda^{2}} + \mathcal{O}\left(\dfrac{v_{H}^4}{\Lambda^4}\right)\,,
\end{align}
where the ingoing gluons have color indices $a,b$, momenta $p_1^{\mu}$ and $p_2^{\nu}$ and are assumed to be on-shell. We denote the metric tensor with (1,-1,-1,-1) signature by $g_{\mu\nu}$.
The Wilson coefficient, computed with \texttt{Matchete} \cite{Fuentes-Martin:2022jrf}, is
\begin{equation}
    \frac{C_{HG}}{\Lambda^2} = \frac{1}{4\pi} \frac{\alpha_{s}(\mu) c_{\lambda\phi}}{48 M_{\text{ex}}^2} \,.
\end{equation}
In the broken phase this terms generates two effective couplings of the Higgs boson with gluons. We recall our normalization for later convenience:
\begin{align}
     & \Delta{\cal L}_{\text{SMEFT}}^{b.p.} \supset  \frac{\alpha_s}{8\pi} \left( c_{ggh}^{\text{SMEFT}} \frac{h}{v}+
c_{gghh}^{\text{SMEFT}}\frac{h^2}{v^2}  \right)\, G^a_{\mu \nu} G^{a,\mu \nu}\;, \quad c_{ggh}^{\text{SMEFT}} = \frac{c_{\lambda \phi}v_{H}^2}{24 M_{\text{ex}}^2}, \quad c_{gghh}^{\text{SMEFT}} = \frac{c_{\lambda \phi}v_{H}^2}{48 M_{\text{ex}}^2} \,.
\end{align}
Analogously to the scalar singlet case, on-shell matching to the HEFT basis has been performed through a diagrammatic approach at the one-loop level. Taking $M_{\text{Loryon}}^2 = M_{\text{ex}}^2 + \frac{c_{\lambda\phi}v_{H}^2}{4}$ or equivalently $M_{\text{Loryon}}^2 = M_{\text{ex}}^2 + \Delta M_{\text{Loryon}}^2$ the effective couplings read:
\begin{align}
    & c_{ggh}=\frac{c_{\lambda\phi} v_{H}^2}{24 M_{\text{Loryon}}^2}= \frac{\Delta M_{\text{Loryon}}^2}{6 M_{\text{Loryon}}^2}, \\
    & c_{gghh}=\frac{c_{\lambda\phi} v_{H}^2}{48 M_{\text{Loryon}}^2}-\frac{(c_{\lambda\phi} v_{H}^2)^2}{96 M_{\text{Loryon}}^4} = \frac{\Delta M_{\text{Loryon}}^2}{12 M_{\text{Loryon}}^2}-\frac{\Delta M_{\text{Loryon}}^4}{6 M_{\text{Loryon}}^4} \label{eq:c_gghh_HEFT}\,.
\end{align}
\begin{comment}
\begin{align}
  & c_{ggh}=\frac{c_{\lambda\phi} v_{H}^2}{6(4M_{\text{ex}}^2+c_{\lambda\phi}v_{H}^2)}+\frac{8 p_1 p_2 c_{\lambda\phi}v_{H}^2}{45(4M_{\text{ex}}^2+c_{\lambda\phi}v_{H}^2)^2} , \\ & c_{gghh}=\frac{c_{\lambda\phi} v_{H}^2}{12(4M_{\text{ex}}^2+c_{\lambda\phi}v_{H}^2)}+\frac{8 p_1 p_2 c_{\lambda\phi}v_{H}^2}{90(4M_{\text{ex}}^2+c_{\lambda\phi}v_{H}^2)^2}-\frac{(c_{\lambda\phi} v_{H}^2)^2}{6 (4M_{\text{ex}}^2+c_{\lambda\phi}v_{H}^2)^2}  \label{eq:c_gghh_HEFT}\,.
\end{align}
\end{comment}
%Terms proportional to $p_1$ and $p_2$, which are the external gluon momenta %\textcolor{red}{LT: maybe belong to  $d_{\chi}=4$ basis \cite{Sun_2023}},
%are not generated by $\mathcal{O}_{HG}$ and should be discarded in our analysis. \\
Similarly to the other cases discussed before, the SMEFT predicts a linear relation among the effective couplings being $c_{gghh}^{\text{SMEFT}}= 1/2 \,c_{ggh}^{\text{SMEFT}}$. The relation is broken in HEFT due to the contribution of the last term in Eq.~\eqref{eq:c_gghh_HEFT}.
In the limit $M_{\text{ex}}^2 \gg c_{\lambda \phi}v_{H}^2$, we see though that $c_{ggh}=c_{ggh}^{\text{SMEFT}}$ and $c_{gghh}=c_{gghh}^{\text{SMEFT}}$, so the SMEFT relation is re-established. Indeed, this is in analogy to the case of a heavy quark loop in gluon fusion: as stated in \cite{Pierce:2006dh} models where the quark gets its whole mass from electroweak symmetry breaking match to an operator 
\begin{equation}
\log\left( \frac{H^{\dagger} H}{v^2} \right)G^a_{\mu\nu}G^{a,\mu\nu}
\end{equation}
while quarks that get mass from an explicit mass term match to $\mathcal{O}_{HG}$.
General models give something in between.
\subsubsection{Results for gluon fusion into a Higgs pair}
\begin{figure}[!t]
\begin{center}
\includegraphics[scale=0.9]{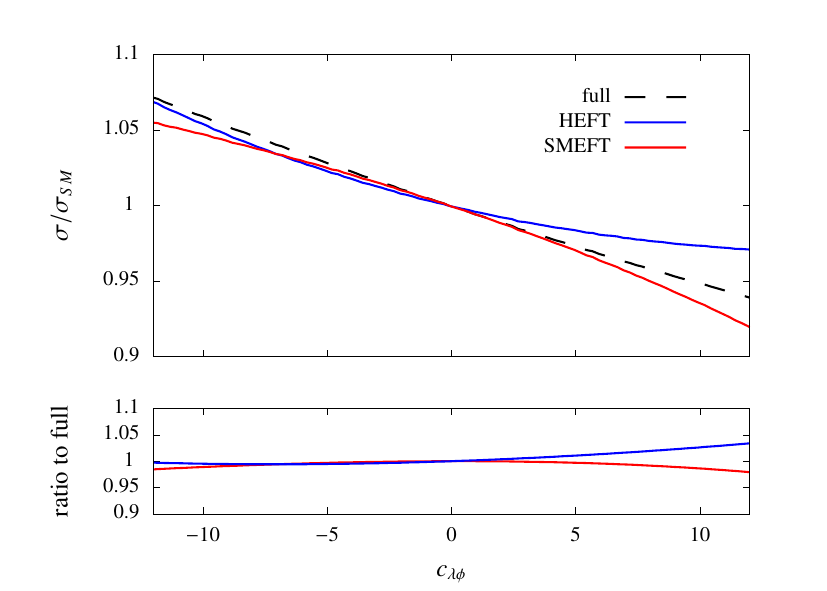}
\caption{\textit{Upper pannel:} Cross section divided by the SM value for the UV model of a colored scalar (black dashed), HEFT (blue) and SMEFT (red) setting $M_{\text{Loryon}}=1\text{ TeV}$. \textit{Lower pannel:} Ratio of SMEFT (red) and HEFT (blue) cross section with respect to the cross section in the UV model. \label{fig:coloredscalar}}
\end{center}
\end{figure}
In Fig.~\ref{fig:coloredscalar} we present the results for this model as a function of $c_{\lambda \phi}$ for a mass of $m_{\omega_1}=1\text{ TeV}$.
We have explicitly checked that when computing all the  matching coefficients at one-loop level in the model,  Higgs measurements do not lead to any relevant bound on $c_{\lambda \phi}$, on which we hence impose by perturbative unitarity $|c_{\lambda \phi}|<4\pi$. As can be inferred from Fig.~\ref{fig:coloredscalar}, the color scalar model is slightly better described by HEFT than by SMEFT. The cross section though changes only little with respect to the SM value even for large $c_{\lambda\phi}$, much below the theoretical uncertainty for the gluon fusion cross section \cite{DiMicco:2019ngk}. 
%%%%%%%%%%%%%%%%%%%%%%%%%%%%%%
\section{Conclusion \label{sec:conclusion}}
In this work, we investigate whether new-physics effects in Higgs pair production require a HEFT description, or whether they can be adequately captured within SMEFT.
Since current experimental searches are performed within the framework of dimension-6 SMEFT and the LO HEFT Lagrangian, we restrict our analysis accordingly.

To address this question, we studied three models that can act as so-called Loryons -- particles that acquire more than half their mass from electroweak symmetry breaking and are thus better suited to HEFT than SMEFT descriptions. These include the scalar singlet extension, the 2HDM, and an extension with a colored scalar. At tree level, the singlet model generates contributions to $C_{H\Box}$ and $C_H$, leading to a general rescaling of Higgs couplings and modifications to the trilinear Higgs self-coupling. The 2HDM additionally generates contributions to $C_{tH}$, modifying the Higgs couplings to top quarks  while keeping the Higgs to vector boson couplings unaffected, and the colored scalar model generates contributions to $C_{HG}$ at one-loop order.

We explicitly demonstrated the conditions under which the relevant HEFT and SMEFT couplings agree. This connection relates to the fact that HEFT assumes large physical masses, whereas SMEFT matching is performed in the unbroken phase as an EFT expansion in a large Lagrangian mass parameter.

Our results explicitly show that for both the singlet and 2HDM, there exists viable parameter space where HEFT provides a demonstrably better description of the UV model than SMEFT. For the colored scalar, however, since we considered loop-induced couplings, the difference in the cross section relative to the SM was sufficiently small to fall below measurement uncertainty.

Finally, we note that our study was restricted to the dominant EFT contributions -- tree-level matching for the singlet and 2HDM, and one-loop matching for the colored scalar. Furthermore, we considered only leading-order HEFT and dimension-6 SMEFT. Future extensions of this study could incorporate loop effects and higher orders in the EFT expansion.

\begin{comment}

taken a first step towards the analysis of SMEFT vs HEFT. We have seen that SMEFT is not predictive with respect to HEFT if one restricts the comparison to the expansion order considered in this analysis.
\begin{itemize}
    \item Comment results: scalar singlet model shows a sizable effect depending on the EFT that is adopted. From the operational point of view integrating out the field in the gauge eigenstates and later breaking the EW symmetry is not equivalent to integrating out the heavy field in mass eigenstate.
    2HDM.....
    Colored scalar model is quite insensitive to the EFT withing the perturbative scenario \textcolor{red}{LT:Could this be expected before? In the end $\omega_1$ is $SU(2)$ singlet...}.  
    
    \item Can the discrepancy of SMEFT matching be fixed by inclusion of higher-order terms (i.e. dim-8 and/or loops)?1-loop matching of the scalar singlet only generates a small set of operators, doable computation in principle, 1-loop corrections of 2HDM may be more complex and large corrections to trilinear higgs self coupling are expected. A loop computation would require a careful treatment of:
    \begin{itemize}
        \item $\gamma_5$ scheme in D dimension.
        \item Mixing and running of WC generated at the matching scale.
    
    \end{itemize}
    
\end{itemize}
\end{comment}
%%%%%%%%%%%%%%%%%%%%%%%%%%%%%%

%%%%%%%%%%%%%%%%%%%
\subsection*{Acknowledgements}
%%%%%%%%%%%%%%%%%%%
RG thanks Ilaria Brivio and Konstantin Schmid for endless HEFT discussions. The work of RG is supported by a STARS@UniPD grant under the acronym ``HiggsPairs'' and in part by the Italian MUR Departments of Excellence grant 2023-2027 ”Quantum Frontiers”.  The authors acknowledge support from the COMETA COST Action CA22130.

%%%%%%%%%%%%%%%%%%%
\appendix
\section{Definition of HEFT coefficients \label{app:couplings}}
In this appendix we collect the relevant quantities for the matching to HEFT.
\subsection{Scalar singlet model}
For the scalar singlet model discussed in section \ref{sec:singlet} we give the analytic expression for the light Higgs boson $h$ self couplings and the couplings of a heavy Higgs boson to two light Higgs bosons in terms of the model parameters:
\begin{align}
   d_1 = & v_{H} \lambda_{H}\cos^3 \theta - \frac{1}{2} \lambda_3 v_S \cos^2\theta \sin \theta - \frac{\mu_4}{2} \cos^2 \theta \sin\theta + \frac{1}{2}\lambda_3 v_{H} \cos\theta \sin^2\theta \notag \\
   &- \lambda_2 v_S \sin^3\theta - \frac{\mu_3}{3} \sin^3\theta\,, \\ 
   d_2 = &  \frac{1}{2} \lambda_3 v_S \cos^3 \theta + \frac{1}{2} \mu_4 \cos^3\theta + 3 \lambda_H v_H \cos^2\theta \sin\theta - \lambda_3 v_H \cos^2\theta \sin\theta \notag \\
   & + 3 v_S \lambda_2 \cos\theta \sin^2\theta + \mu_3 \cos\theta \sin^2\theta 
    - \mu_4 \cos\theta \sin^2\theta - \lambda_3 v_S \cos\theta \sin^2\theta  \notag \\
    & + \frac{1}{2} \lambda_3 v_H \sin^3\theta \,. 
\end{align}
\subsection{Two Higgs doublet model}
For the 2HDM, we take the Higgs self-couplings necessary for the HEFT matching in Eqs.~\eqref{eq:chhh2HDM} and \eqref{eq:c2t2HDM} from Ref.~\cite{Arco:2025pgx}:
\begin{align}
d_1^{\text{2HDM}} =&  \frac{3}{2v^2}\left[s_{\beta-\alpha}^3 m_h^2+s_{\beta-\alpha} c_{\beta-\alpha}^2 \left(3 m_h^2-2 M^2\right)  + 2 c_{\beta-\alpha}^3 \cot2\beta \left(m_h^2-M^2\right)\right], 
	\label{eq:d1THDM} \\
d_2^{\text{2HDM}} =& \frac{-c_{\beta-\alpha} }{v^2}  \left[s_{\beta-\alpha}^2 \left(2 m_h^2+m_H^2-4 M^2\right)  +  2 s_{\beta-\alpha} c_{\beta-\alpha} \cot2\beta 
\left(2 m_h^2+m_H^2-3 M^2\right) \nonumber \right. \\  & - \left. c_{\beta-\alpha}^2 \left(2 m_h^2+m_H^2-2 M^2\right)\right]\,.\label{eq:d2THDM}
\end{align}
%%%%%%%%%%%%%%%%%%%%%%%%%%%%%%%%%%%%%%%%%%%%%%%%%%%%%%%%%%%%%%%%%%%%%%%%%%%%%%%%%%%%%%%%%%%%%singlet plots vs mixing angle%%%%%%%%%%%%%%%%%%%%%%%%%%%%%%%%%%%%%%%%%%%%%%%%%%%%%%%%%%%%%%%%%%%%%%%%%%%%%%%%%%%%%%%
\section{Singlet Model: mixing angle dependece}
\begin{figure}[H]
    \centering
%    % Riga superiore
%    \includegraphics[width=0.45\textwidth]{ggF_singlet_xs_mix_angle/SMEFT_UV_lowvs.pdf} \hfill
 %   \includegraphics[width=0.45\textwidth]{ggF_singlet_xs_mix_angle/HEFT_UV_lowvs.pdf}
    
%    \vspace{0.5cm} % spazio verticale tra le righe
    
    % Riga inferiore
    \includegraphics[width=0.45\textwidth]{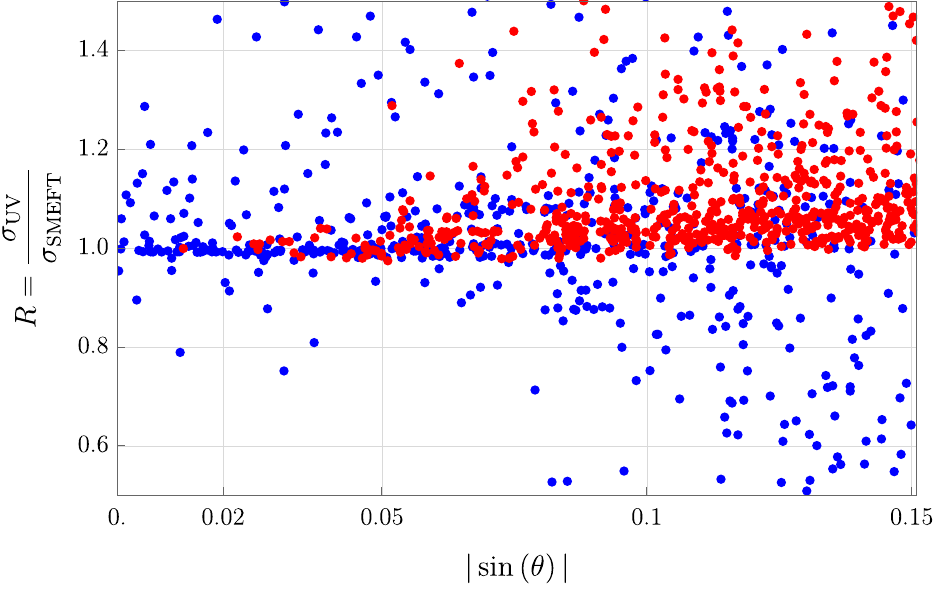} \hfill
    \includegraphics[width=0.45\textwidth]{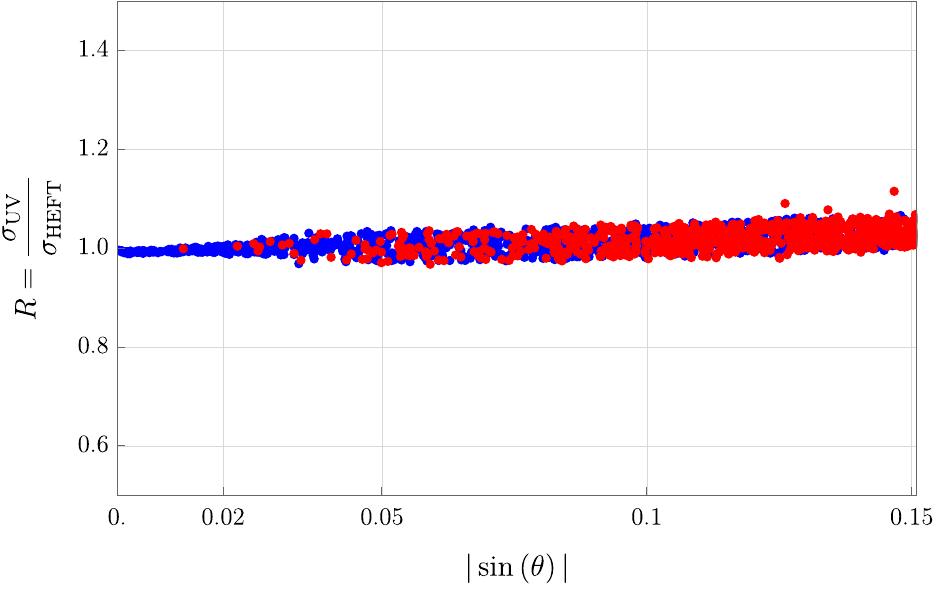}
    
    \caption{Same as Fig.~\ref{fig::xs for ggF in the scalar singlet}, but shown in terms of $|\sin\theta|$. %\RG{is the upper one needed? does the lower panel not show the same but more?} %
    \label{fig:singletfunctionmixangle}
    }
\end{figure}

Here we show Fig.~\ref{fig:singletfunctionmixangle} already presented as Fig.~\ref{fig::xs for ggF in the scalar singlet} but as  function of the mixing angle $\sin \theta$. 

%%%%%%%%%%%%%%%%%%%
\section{Feynman Rules for the Colored Scalar Model\label{app:Feynman Rules}}
Here we present the Feynman rules needed for the computation of the cross sections for the colored scalar model. They were obtained with \texttt{FeynRules} \cite{Alloul:2013bka}. %The indices $a_i$ take values from $a_i= 1,2,...,8$ while $m_i = 1,2,3$. %
$T^a$ are the colour matrices with a generic matrix element written as $T^{a}_{m_i m_j}$.
\newline
\newline
\vspace{1 cm}
% 1) (G, omega1, omega1^\dagger)
\DiagWithRule{%
  \begin{tikzpicture}
    \begin{feynman}
      \vertex (G) at (-3.2,0) {$G^{a_1}(p_1)_{\mu_1}$};
      \vertex (v) at (0,0);
      \vertex (o1) at (2,1) {$\omega_{1}(p_2)_{m_2}$};
      \vertex (o1d) at (2,-1) {$\omega_{1}^\dagger(p_3)_{m_3}$};
      \diagram*{
        (G) -- [gluon] (v),
        (v) -- [scalar] (o1),
        (v) -- [scalar] (o1d),
      };
    \end{feynman}
  \end{tikzpicture}
}{%
  i\,g_s\!\left(p_2 -p_3 \right) ^{\mu_1} \,T_{\,\,m_3 \,m_2}^{\,a_1}
}

% 3) (G, G, omega1, omega1^\dagger)
\DiagWithRule{%
  \begin{tikzpicture}
    \begin{feynman}
      \vertex (G1) at (-3.2,1) {$G^{a_1}(p_1)_{\mu_1}$};
      \vertex (G2) at (-3.2,-1) {$G^{a_2}(p_2)_{\mu_2}$};
      \vertex (v) at (0,0);
      \vertex (o1) at (3,1) {$\omega_{1}(p_3)_{m_3}$};
      \vertex (o1d) at (3,-1) {$\omega_{1}^\dagger(p_4)_{m_4}$};
      \diagram*{
        (G1) -- [gluon] (v),
        (G2) -- [gluon] (v),
        (v) -- [scalar] (o1),
        (v) -- [scalar] (o1d),
      };
    \end{feynman}
  \end{tikzpicture}
}{%
  i\,g_s^2\,g_{\mu_1 \mu_2}\!\left( T^{a_1}_{m_4 , \, b}T^{a_2}_{b\, m_3} + T^{a_1}_{b, \, m_3}T^{a_2}_{m4, b} \right)
}

% 8) (H, H, omega1, omega1^\dagger)
\DiagWithRule{%
  \begin{tikzpicture}
    \begin{feynman}
      \vertex (H1) at (-3.2,1) {$H(p_1)$};
      \vertex (H2) at (-3.2,-1) {$H(p_2)$};
      \vertex (v) at (0,0);
      \vertex (o1) at (3,1) {$\omega_{1}(p_3)_{m_3}$};
      \vertex (o1d) at (3,-1) {$\omega_{1}^\dagger(p_4)_{m_4}$};
      \diagram*{
        (H1) -- [scalar] (v),
        (H2) -- [scalar] (v),
        (v) -- [scalar] (o1),
        (v) -- [scalar] (o1d),
      };
    \end{feynman}
  \end{tikzpicture}
}{%
  -\dfrac{1}{2} i\,c_{\lambda h}\,\delta_{m_3 m_4}
}

% 9) (H, omega1, omega1^\dagger)
\DiagWithRule{%
  \begin{tikzpicture}
    \begin{feynman}
      \vertex (H) at (-3.2,0) {$H(p_1)$};
      \vertex (v) at (0,0);
      \vertex (o1) at (2,1) {$\omega_{1}(p_2)_{m_2}$};
      \vertex (o1d) at (2,-1) {$\omega_{1}^\dagger(p_3)_{m_3}$};
      \diagram*{
        (H) -- [scalar] (v),
        (v) -- [scalar] (o1),
        (v) -- [scalar] (o1d),
      };
    \end{feynman}
  \end{tikzpicture}
}{%
  -\dfrac{1}{2} i\ v_{H}\,c_{\lambda h}\,\delta_{m_2 m_3}
}

%%%%%%%%%%%%%%%%%%%

\clearpage
\newpage

%%%%%%%%%%%%%%%%%%%
\bibliographystyle{utphys.bst}
\bibliography{bibliography}

@article{ATLAS:2012yve,
    author = "Aad, Georges and others",
    collaboration = "ATLAS",
    title = "{Observation of a new particle in the search for the Standard Model Higgs boson with the ATLAS detector at the LHC}",
    eprint = "1207.7214",
    archivePrefix = "arXiv",
    primaryClass = "hep-ex",
    reportNumber = "CERN-PH-EP-2012-218",
    doi = "10.1016/j.physletb.2012.08.020",
    journal = "Phys. Lett. B",
    volume = "716",
    pages = "1--29",
    year = "2012"
}

@article{CMS:2012qbp,
    author = "Chatrchyan, Serguei and others",
    collaboration = "CMS",
    title = "{Observation of a New Boson at a Mass of 125 GeV with the CMS Experiment at the LHC}",
    eprint = "1207.7235",
    archivePrefix = "arXiv",
    primaryClass = "hep-ex",
    reportNumber = "CMS-HIG-12-028, CERN-PH-EP-2012-220",
    doi = "10.1016/j.physletb.2012.08.021",
    journal = "Phys. Lett. B",
    volume = "716",
    pages = "30--61",
    year = "2012"
}

@article{ATLAS:2022vkf,
    author = "Aad, Georges and others",
    collaboration = "ATLAS",
    title = "{A detailed map of Higgs boson interactions by the ATLAS experiment ten years after the discovery}",
    eprint = "2207.00092",
    archivePrefix = "arXiv",
    primaryClass = "hep-ex",
    reportNumber = "CERN-EP-2022-057",
    doi = "10.1038/s41586-022-04893-w",
    journal = "Nature",
    volume = "607",
    number = "7917",
    pages = "52--59",
    year = "2022",
    note = "[Erratum: Nature 612, E24 (2022)]"
}

@article{CMS:2022dwd,
    author = "Tumasyan, Armen and others",
    collaboration = "CMS",
    title = "{A portrait of the Higgs boson by the CMS experiment ten years after the discovery.}",
    eprint = "2207.00043",
    archivePrefix = "arXiv",
    primaryClass = "hep-ex",
    reportNumber = "CMS-HIG-22-001, CERN-EP-2022-039",
    doi = "10.1038/s41586-022-04892-x",
    journal = "Nature",
    volume = "607",
    number = "7917",
    pages = "60--68",
    year = "2022"
}

@article{Djouadi:1999rca,
    author = "Djouadi, A. and Kilian, W. and Muhlleitner, M. and Zerwas, P. M.",
    title = "{Production of neutral Higgs boson pairs at LHC}",
    eprint = "hep-ph/9904287",
    archivePrefix = "arXiv",
    reportNumber = "DESY-99-033, TTP-99-17, PM-99-21",
    doi = "10.1007/s100529900083",
    journal = "Eur. Phys. J. C",
    volume = "10",
    pages = "45--49",
    year = "1999"
}

@article{Baglio:2012np,
    author = {Baglio, J. and Djouadi, A. and Gr{\"o}ber, R. and M{\"u}hlleitner, M. M. and Quevillon, J. and Spira, M.},
    title = "{The measurement of the Higgs self-coupling at the LHC: theoretical status}",
    eprint = "1212.5581",
    archivePrefix = "arXiv",
    primaryClass = "hep-ph",
    reportNumber = "KA-TP-44-2012, SFB-CPP-12-102, LPT-ORSAY-12-124, PSI-PR-12-10",
    doi = "10.1007/JHEP04(2013)151",
    journal = "JHEP",
    volume = "04",
    pages = "151",
    year = "2013"
}

@article{Dolan:2012rv,
    author = "Dolan, Matthew J. and Englert, Christoph and Spannowsky, Michael",
    title = "{Higgs self-coupling measurements at the LHC}",
    eprint = "1206.5001",
    archivePrefix = "arXiv",
    primaryClass = "hep-ph",
    reportNumber = "IPPP-12-43, DCPT-12-86",
    doi = "10.1007/JHEP10(2012)112",
    journal = "JHEP",
    volume = "10",
    pages = "112",
    year = "2012"
}

@article{BUCHMULLER1986621,
title = {Effective lagrangian analysis of new interactions and flavour conservation},
journal = {Nuclear Physics B},
volume = {268},
number = {3},
pages = {621-653},
year = {1986},
issn = {0550-3213},
doi = {https://doi.org/10.1016/0550-3213(86)90262-2},
author = {W. Buchm{\"u}ller and D. Wyler}
}

@article{dim6smeft,
    author = "Grzadkowski, B. and Iskrzynski, M. and Misiak, M. and Rosiek, J.",
    title = "{Dimension-Six Terms in the Standard Model Lagrangian}",
    eprint = "1008.4884",
    archivePrefix = "arXiv",
    primaryClass = "hep-ph",
    reportNumber = "IFT-9-2010, TTP10-35",
    doi = "10.1007/JHEP10(2010)085",
    journal = "JHEP",
    volume = "10",
    pages = "085",
    year = "2010"
}

@article{Brivio:2017vri,
    author = "Brivio, Ilaria and Trott, Michael",
    title = "{The Standard Model as an Effective Field Theory}",
    eprint = "1706.08945",
    archivePrefix = "arXiv",
    primaryClass = "hep-ph",
    doi = "10.1016/j.physrep.2018.11.002",
    journal = "Phys. Rept.",
    volume = "793",
    pages = "1--98",
    year = "2019"
}

@article{Feruglio:1992wf,
    author = "Feruglio, F.",
    title = "{The Chiral approach to the electroweak interactions}",
    eprint = "hep-ph/9301281",
    archivePrefix = "arXiv",
    reportNumber = "DFPD-92-TH-50",
    doi = "10.1142/S0217751X93001946",
    journal = "Int. J. Mod. Phys. A",
    volume = "8",
    pages = "4937--4972",
    year = "1993"
}

@article{Buchalla:2012qq,
    author = "Buchalla, Gerhard and Cata, Oscar",
    title = "{Effective Theory of a Dynamically Broken Electroweak Standard Model at NLO}",
    eprint = "1203.6510",
    archivePrefix = "arXiv",
    primaryClass = "hep-ph",
    reportNumber = "LMU-ASC-19-12, FLAVOUR-267104-ERC-9, LMU-ASC\textasciitilde{}19-12",
    doi = "10.1007/JHEP07(2012)101",
    journal = "JHEP",
    volume = "07",
    pages = "101",
    year = "2012"
}

@article{Alonso:2012px,
    author = "Alonso, R. and Gavela, M. B. and Merlo, L. and Rigolin, S. and Yepes, J.",
    title = "{The Effective Chiral Lagrangian for a Light Dynamical ''Higgs Particle''}",
    eprint = "1212.3305",
    archivePrefix = "arXiv",
    primaryClass = "hep-ph",
    reportNumber = "FTUAM-12-115, IFT-UAM-CSIC-12-113, CERN-PH-TH-2012-335, DFPD-2012-TH-23",
    doi = "10.1016/j.physletb.2013.04.037",
    journal = "Phys. Lett. B",
    volume = "722",
    pages = "330--335",
    year = "2013",
    note = "[Erratum: Phys.Lett.B 726, 926 (2013)]"
}

@article{Brivio:2013pma,
    author = "Brivio, I. and Corbett, T. and \'Eboli, O. J. P. and Gavela, M. B. and Gonzalez-Fraile, J. and Gonzalez-Garcia, M. C. and Merlo, L. and Rigolin, S.",
    title = "{Disentangling a dynamical Higgs}",
    eprint = "1311.1823",
    archivePrefix = "arXiv",
    primaryClass = "hep-ph",
    reportNumber = "FTUAM-13-32, IFT-UAM-CSIC-13-119, YITP-SB-13-33, DFPD-2013-TH-20",
    doi = "10.1007/JHEP03(2014)024",
    journal = "JHEP",
    volume = "03",
    pages = "024",
    year = "2014"
}

@article{Buchalla:2015wfa,
    author = "Buchalla, G. and Cata, O. and Celis, A. and Krause, C.",
    title = "{Note on Anomalous Higgs-Boson Couplings in Effective Field Theory}",
    eprint = "1504.01707",
    archivePrefix = "arXiv",
    primaryClass = "hep-ph",
    reportNumber = "LMU-ASC-19-15",
    doi = "10.1016/j.physletb.2015.09.027",
    journal = "Phys. Lett. B",
    volume = "750",
    pages = "298--301",
    year = "2015"
}

@article{Burgess:1999ha,
    author = "Burgess, C. P. and Matias, J. and Pospelov, M.",
    title = "{A Higgs or not a Higgs? What to do if you discover a new scalar particle}",
    eprint = "hep-ph/9912459",
    archivePrefix = "arXiv",
    reportNumber = "CERN-TH-99-311, MCGILL-99-33, TPI-MINN-99-53, UMN-TH-1828-99",
    doi = "10.1142/S0217751X02009813",
    journal = "Int. J. Mod. Phys. A",
    volume = "17",
    pages = "1841--1918",
    year = "2002"
}

@article{CMS:2022gjd,
    author = "Tumasyan, Armen and others",
    collaboration = "CMS",
    title = "{Search for Nonresonant Pair Production of Highly Energetic Higgs Bosons Decaying to Bottom Quarks}",
    eprint = "2205.06667",
    archivePrefix = "arXiv",
    primaryClass = "hep-ex",
    reportNumber = "CMS-B2G-22-003, CERN-EP-2022-090",
    doi = "10.1103/PhysRevLett.131.041803",
    journal = "Phys. Rev. Lett.",
    volume = "131",
    number = "4",
    pages = "041803",
    year = "2023"
}

@article{ATLAS:2022fxe,
    collaboration = "ATLAS",
    title = "{HEFT interpretations of Higgs boson pair searches in $b\bar{b} \gamma \gamma$ and $b\bar{b}\tau \tau$ final states and of their combination in ATLAS}",
    note = "ATL-PHYS-PUB-2022-019",
    year = "2022"
}

@article{Banta:2021dek,
    author = "Banta, Ian and Cohen, Timothy and Craig, Nathaniel and Lu, Xiaochuan and Sutherland, Dave",
    title = "{Non-decoupling new particles}",
    eprint = "2110.02967",
    archivePrefix = "arXiv",
    primaryClass = "hep-ph",
    doi = "10.1007/JHEP02(2022)029",
    journal = "JHEP",
    volume = "02",
    pages = "029",
    year = "2022"
}

@article{Cohen:2020xca,
    author = "Cohen, Timothy and Craig, Nathaniel and Lu, Xiaochuan and Sutherland, Dave",
    title = "{Is SMEFT Enough?}",
    eprint = "2008.08597",
    archivePrefix = "arXiv",
    primaryClass = "hep-ph",
    doi = "10.1007/JHEP03(2021)237",
    journal = "JHEP",
    volume = "03",
    pages = "237",
    year = "2021"
}

@article{deBlas:2017xtg,
    author = "de Blas, J. and Criado, J. C. and Perez-Victoria, M. and Santiago, J.",
    title = "{Effective description of general extensions of the Standard Model: the complete tree-level dictionary}",
    eprint = "1711.10391",
    archivePrefix = "arXiv",
    primaryClass = "hep-ph",
    reportNumber = "CERN-TH-2017-251",
    doi = "10.1007/JHEP03(2018)109",
    journal = "JHEP",
    volume = "03",
    pages = "109",
    year = "2018"
}

@article{Contino:2010mh,
    author = "Contino, Roberto and Grojean, Christophe and Moretti, Mauro and Piccinini, Fulvio and Rattazzi, Riccardo",
    title = "{Strong Double Higgs Production at the LHC}",
    eprint = "1002.1011",
    archivePrefix = "arXiv",
    primaryClass = "hep-ph",
    reportNumber = "CERN-PH-TH-2009-036",
    doi = "10.1007/JHEP05(2010)089",
    journal = "JHEP",
    volume = "05",
    pages = "089",
    year = "2010"
}

@article{Contino:2013kra,
    author = "Contino, Roberto and Ghezzi, Margherita and Grojean, Christophe and Muhlleitner, Margarete and Spira, Michael",
    title = "{Effective Lagrangian for a light Higgs-like scalar}",
    eprint = "1303.3876",
    archivePrefix = "arXiv",
    primaryClass = "hep-ph",
    reportNumber = "CERN-PH-TH-2013-047, KA-TP-06-2013, PSI-PR-13-04",
    doi = "10.1007/JHEP07(2013)035",
    journal = "JHEP",
    volume = "07",
    pages = "035",
    year = "2013"
}

@article{Dawson:1998py,
    author = "Dawson, S. and Dittmaier, S. and Spira, M.",
    title = "{Neutral Higgs boson pair production at hadron colliders: QCD corrections}",
    eprint = "hep-ph/9805244",
    archivePrefix = "arXiv",
    reportNumber = "DESY-98-028, BNL-HET-98-15, CERN-TH-98-135",
    doi = "10.1103/PhysRevD.58.115012",
    journal = "Phys. Rev. D",
    volume = "58",
    pages = "115012",
    year = "1998"
}

@article{Baglio:2018lrj,
    author = {Baglio, Julien and Campanario, Francisco and Glaus, Seraina and M\"uhlleitner, Margarete and Spira, Michael and Streicher, Juraj},
    title = "{Gluon fusion into Higgs pairs at NLO QCD and the top mass scheme}",
    eprint = "1811.05692",
    archivePrefix = "arXiv",
    primaryClass = "hep-ph",
    reportNumber = "FTUV-18-1113, IFIC/18-55, KA-TP-32-2018, PSI-PR-18-14",
    doi = "10.1140/epjc/s10052-019-6973-3",
    journal = "Eur. Phys. J. C",
    volume = "79",
    number = "6",
    pages = "459",
    year = "2019"
}

@article{Borowka:2016ypz,
    author = "Borowka, S. and Greiner, N. and Heinrich, G. and Jones, S. P. and Kerner, M. and Schlenk, J. and Zirke, T.",
    title = "{Full top quark mass dependence in Higgs boson pair production at NLO}",
    eprint = "1608.04798",
    archivePrefix = "arXiv",
    primaryClass = "hep-ph",
    reportNumber = "MPP-2016-261, ZU-TH-31-16",
    doi = "10.1007/JHEP10(2016)107",
    journal = "JHEP",
    volume = "10",
    pages = "107",
    year = "2016"
}

@article{Borowka:2016ehy,
    author = "Borowka, S. and Greiner, N. and Heinrich, G. and Jones, S. P. and Kerner, M. and Schlenk, J. and Schubert, U. and Zirke, T.",
    title = "{Higgs Boson Pair Production in Gluon Fusion at Next-to-Leading Order with Full Top-Quark Mass Dependence}",
    eprint = "1604.06447",
    archivePrefix = "arXiv",
    primaryClass = "hep-ph",
    reportNumber = "MPP-2016-80, NSF-KITP-16-040, ZH-TH-14-16",
    doi = "10.1103/PhysRevLett.117.079901",
    journal = "Phys. Rev. Lett.",
    volume = "117",
    number = "1",
    pages = "012001",
    year = "2016",
    note = "[Erratum: Phys.Rev.Lett. 117, 079901 (2016)]"
}

@article{Grazzini:2018bsd,
    author = "Grazzini, Massimiliano and Heinrich, Gudrun and Jones, Stephen and Kallweit, Stefan and Kerner, Matthias and Lindert, Jonas M. and Mazzitelli, Javier",
    title = "{Higgs boson pair production at NNLO with top quark mass effects}",
    eprint = "1803.02463",
    archivePrefix = "arXiv",
    primaryClass = "hep-ph",
    reportNumber = "CERN-TH-2018-044, IPPP/18/15, MPP-2018-30, ZU-TH 10/18, IPPP-18-15, ZU-TH-10-18",
    doi = "10.1007/JHEP05(2018)059",
    journal = "JHEP",
    volume = "05",
    pages = "059",
    year = "2018"
}

@article{AH:2022elh,
    author = "A H, Ajjath and Shao, Hua-Sheng",
    title = "{N$^{3}$LO+N$^{3}$LL QCD improved Higgs pair cross sections}",
    eprint = "2209.03914",
    archivePrefix = "arXiv",
    primaryClass = "hep-ph",
    doi = "10.1007/JHEP02(2023)067",
    journal = "JHEP",
    volume = "02",
    pages = "067",
    year = "2023"
}

@article{Dreyer:2018qbw,
    author = "Dreyer, Fr\'ed\'eric A. and Karlberg, Alexander",
    title = "{Vector-Boson Fusion Higgs Pair Production at N$^3$LO}",
    eprint = "1811.07906",
    archivePrefix = "arXiv",
    primaryClass = "hep-ph",
    reportNumber = "OUTP-18-09P, ZU-TH 38/18",
    doi = "10.1103/PhysRevD.98.114016",
    journal = "Phys. Rev. D",
    volume = "98",
    number = "11",
    pages = "114016",
    year = "2018"
}

@article{Ling:2014sne,
    author = "Ling, Liu Sheng and Zhang, Ren You and Ma, Wen-Gan and Guo, Lei and Li, Wei Hua and Li, Xiao Zhou",
    title = "{NNLO QCD corrections to Higgs pair production via vector boson fusion at hadron colliders}",
    eprint = "1401.7754",
    archivePrefix = "arXiv",
    primaryClass = "hep-ph",
    doi = "10.1103/PhysRevD.89.073001",
    journal = "Phys. Rev. D",
    volume = "89",
    number = "7",
    pages = "073001",
    year = "2014"
}

@article{Dreyer:2020xaj,
    author = "Dreyer, Fr\'ed\'eric A. and Karlberg, Alexander and Lang, Jean-Nicolas and Pellen, Mathieu",
    title = "{Precise predictions for double-Higgs production via vector-boson fusion}",
    eprint = "2005.13341",
    archivePrefix = "arXiv",
    primaryClass = "hep-ph",
    doi = "10.1140/epjc/s10052-020-08610-7",
    journal = "Eur. Phys. J. C",
    volume = "80",
    number = "11",
    pages = "1037",
    year = "2020"
}

@article{Alasfar:2023xpc,
    author = "Alasfar, Lina and others",
    title = "{Effective Field Theory descriptions of Higgs boson pair production}",
    eprint = "2304.01968",
    archivePrefix = "arXiv",
    primaryClass = "hep-ph",
    reportNumber = "KA-TP-04-2023, P3H-23-020, OUTP-23-03P, LHCHWG-2022-004",
    doi = "10.21468/SciPostPhysCommRep.2",
    journal = "SciPost Phys. Comm. Rep.",
    volume = "2024",
    pages = "2",
    year = "2024"
}

@article{Arzt:1994gp,
    author = "Arzt, C. and Einhorn, M. B. and Wudka, J.",
    title = "{Patterns of deviation from the standard model}",
    eprint = "hep-ph/9405214",
    archivePrefix = "arXiv",
    reportNumber = "UM-TH-94-15, UCRHEP-125, CALT-68-1932",
    doi = "10.1016/0550-3213(94)00336-D",
    journal = "Nucl. Phys. B",
    volume = "433",
    pages = "41--66",
    year = "1995"
}

@article{DiNoi:2023ygk,
    author = {Di Noi, Stefano and Gr{\"o}ber, Ramona and Heinrich, Gudrun and Lang, Jannis and Vitti, Marco},
    title = "{{\ensuremath{\gamma}}5 schemes and the interplay of SMEFT operators in the Higgs-gluon coupling}",
    eprint = "2310.18221",
    archivePrefix = "arXiv",
    primaryClass = "hep-ph",
    reportNumber = "P3H-23-080, KA-TP-22-2023, TTP23-054",
    doi = "10.1103/PhysRevD.109.095024",
    journal = "Phys. Rev. D",
    volume = "109",
    number = "9",
    pages = "095024",
    year = "2024"
}

@article{Alasfar:2019pmn,
    author = {Alasfar, Lina and Corral Lopez, Roberto and Gr\"ober, Ramona},
    title = "{Probing Higgs couplings to light quarks via Higgs pair production}",
    eprint = "1909.05279",
    archivePrefix = "arXiv",
    primaryClass = "hep-ph",
    reportNumber = "HU-EP-19/25",
    doi = "10.1007/JHEP11(2019)088",
    journal = "JHEP",
    volume = "11",
    pages = "088",
    year = "2019"
}

@article{Alasfar:2022vqw,
    author = {Alasfar, Lina and Gr\"ober, Ramona and Grojean, Christophe and Paul, Ayan and Qian, Zhuoni},
    title = "{Machine learning the trilinear and light-quark Yukawa couplings from Higgs pair kinematic shapes}",
    eprint = "2207.04157",
    archivePrefix = "arXiv",
    primaryClass = "hep-ph",
    reportNumber = "DESY 22-085 | HU-EP-21/34-RTG, DESY 22-085, HU-EP-21/34-RTG",
    doi = "10.1007/JHEP11(2022)045",
    journal = "JHEP",
    volume = "11",
    pages = "045",
    year = "2022"
}

@article{Celada:2023oji,
    author = {Celada, Eugenia and Han, Tao and Kilian, Wolfgang and Kreher, Nils and Ma, Yang and Maltoni, Fabio and Pagani, Davide and Reuter, J{\"u}rgen and Striegl, Tobias and Xie, Keping},
    title = "{Probing Higgs-muon interactions at a multi-TeV muon collider}",
    eprint = "2312.13082",
    archivePrefix = "arXiv",
    primaryClass = "hep-ph",
    reportNumber = "DESY 23-222, PITT-PACC-2325, SI-HEP-2023-33, P3H-23-103,
  IRMP-CP3-23-74, MSUHEP-23-034, COMETA-2023-04",
    doi = "10.1007/JHEP08(2024)021",
    journal = "JHEP",
    volume = "08",
    pages = "021",
    year = "2024"
}

@article{Han:2023njx,
    author = "Han, Tao and Liu, Da and Low, Ian and Wang, Xing",
    title = "{Electroweak scattering at the muon shot}",
    eprint = "2312.07670",
    archivePrefix = "arXiv",
    primaryClass = "hep-ph",
    reportNumber = "PITT-PACC-2324",
    doi = "10.1103/PhysRevD.110.013005",
    journal = "Phys. Rev. D",
    volume = "110",
    number = "1",
    pages = "013005",
    year = "2024"
}

@article{Falkowski:2019tft,
    author = "Falkowski, Adam and Rattazzi, Riccardo",
    title = "{Which EFT}",
    eprint = "1902.05936",
    archivePrefix = "arXiv",
    primaryClass = "hep-ph",
    reportNumber = "LPT Orsay 19-05",
    doi = "10.1007/JHEP10(2019)255",
    journal = "JHEP",
    volume = "10",
    pages = "255",
    year = "2019"
}

@article{Alonso:2016oah,
    author = "Alonso, Rodrigo and Jenkins, Elizabeth E. and Manohar, Aneesh V.",
    title = "{Geometry of the Scalar Sector}",
    eprint = "1605.03602",
    archivePrefix = "arXiv",
    primaryClass = "hep-ph",
    reportNumber = "CERN-TH-2016-116",
    doi = "10.1007/JHEP08(2016)101",
    journal = "JHEP",
    volume = "08",
    pages = "101",
    year = "2016"
}

@article{Alonso:2016btr,
    author = "Alonso, Rodrigo and Jenkins, Elizabeth E. and Manohar, Aneesh V.",
    title = "{Sigma Models with Negative Curvature}",
    eprint = "1602.00706",
    archivePrefix = "arXiv",
    primaryClass = "hep-ph",
    reportNumber = "CERN-TH-2016-024",
    doi = "10.1016/j.physletb.2016.03.032",
    journal = "Phys. Lett. B",
    volume = "756",
    pages = "358--364",
    year = "2016"
}

@article{Alonso:2015fsp,
    author = "Alonso, Rodrigo and Jenkins, Elizabeth E. and Manohar, Aneesh V.",
    title = "{A Geometric Formulation of Higgs Effective Field Theory: Measuring the Curvature of Scalar Field Space}",
    eprint = "1511.00724",
    archivePrefix = "arXiv",
    primaryClass = "hep-ph",
    reportNumber = "CERN-PH-TH-2015-257",
    doi = "10.1016/j.physletb.2016.01.041",
    journal = "Phys. Lett. B",
    volume = "754",
    pages = "335--342",
    year = "2016"
}

@article{Delgado:2023ynh,
    author = "Delgado, Rafael L. and G{\'o}mez-Ambrosio, Raquel and Mart{\'\i}nez-Mart{\'\i}n, Javier and Salas-Bern{\'a}rdez, Alexandre and Sanz-Cillero, Juan J.",
    title = "{Production of two, three, and four Higgs bosons: where SMEFT and HEFT depart}",
    eprint = "2311.04280",
    archivePrefix = "arXiv",
    primaryClass = "hep-ph",
    reportNumber = "IPARCOS-UCM-23-123",
    doi = "10.1007/JHEP03(2024)037",
    journal = "JHEP",
    volume = "03",
    pages = "037",
    year = "2024"
}

@article{Gomez-Ambrosio:2022why,
    author = "G\'omez-Ambrosio, Raquel and Llanes-Estrada, Felipe J. and Salas-Bern\'ardez, Alexandre and Sanz-Cillero, Juan J.",
    title = "{SMEFT is falsifiable through multi-Higgs measurements (even in the absence of new light particles)}",
    eprint = "2207.09848",
    archivePrefix = "arXiv",
    primaryClass = "hep-ph",
    doi = "10.1088/1572-9494/ace95e",
    journal = "Commun. Theor. Phys.",
    volume = "75",
    number = "9",
    pages = "095202",
    year = "2023"
}

@article{Gomez-Ambrosio:2022qsi,
    author = "G\'omez-Ambrosio, Raquel and Llanes-Estrada, Felipe J. and Salas-Bern\'ardez, Alexandre and Sanz-Cillero, Juan J.",
    title = "{Distinguishing electroweak EFTs with $W_LW_L \to n\times h$}",
    eprint = "2204.01763",
    archivePrefix = "arXiv",
    primaryClass = "hep-ph",
    doi = "10.1103/PhysRevD.106.053004",
    journal = "Phys. Rev. D",
    volume = "106",
    number = "5",
    pages = "053004",
    year = "2022"
}

@article{Pierce:2006dh,
    author = "Pierce, Aaron and Thaler, Jesse and Wang, Lian-Tao",
    title = "{Disentangling Dimension Six Operators through Di-Higgs Boson Production}",
    eprint = "hep-ph/0609049",
    archivePrefix = "arXiv",
    reportNumber = "HUTP-06-A0037",
    doi = "10.1088/1126-6708/2007/05/070",
    journal = "JHEP",
    volume = "05",
    pages = "070",
    year = "2007"
}

@article{Fuentes-Martin:2022jrf,
    author = {Fuentes-Mart\'\i{}n, Javier and K\"onig, Matthias and Pag\`es, Julie and Thomsen, Anders Eller and Wilsch, Felix},
    title = "{A proof of concept for matchete: an automated tool for matching effective theories}",
    eprint = "2212.04510",
    archivePrefix = "arXiv",
    primaryClass = "hep-ph",
    reportNumber = "MITP-22-105, TUM-HEP-1443/22, ZU-TH-58/22",
    doi = "10.1140/epjc/s10052-023-11726-1",
    journal = "Eur. Phys. J. C",
    volume = "83",
    number = "7",
    pages = "662",
    year = "2023"
}

@article{Goodsell:2018tti,
    author = "Goodsell, Mark D. and Staub, Florian",
    title = "{Unitarity constraints on general scalar couplings with SARAH}",
    eprint = "1805.07306",
    archivePrefix = "arXiv",
    primaryClass = "hep-ph",
    reportNumber = "KA-TP-10-2018",
    doi = "10.1140/epjc/s10052-018-6127-z",
    journal = "Eur. Phys. J. C",
    volume = "78",
    number = "8",
    pages = "649",
    year = "2018"
}

@article{Lopez-Val:2014jva,
    author = "L\'opez-Val, D. and Robens, T.",
    title = "{\ensuremath{\Delta}r and the W-boson mass in the singlet extension of the standard model}",
    eprint = "1406.1043",
    archivePrefix = "arXiv",
    primaryClass = "hep-ph",
    doi = "10.1103/PhysRevD.90.114018",
    journal = "Phys. Rev. D",
    volume = "90",
    pages = "114018",
    year = "2014"
}

@article{Haisch:2020ahr,
    author = "Haisch, Ulrich and Ruhdorfer, Maximilian and Salvioni, Ennio and Venturini, Elena and Weiler, Andreas",
    title = "{Singlet night in Feynman-ville: one-loop matching of a real scalar}",
    eprint = "2003.05936",
    archivePrefix = "arXiv",
    primaryClass = "hep-ph",
    reportNumber = "CERN-TH-2020-038, TUM-HEP-1254-20",
    doi = "10.1007/JHEP04(2020)164",
    journal = "JHEP",
    volume = "04",
    pages = "164",
    year = "2020",
    note = "[Erratum: JHEP 07, 066 (2020)]"
}

@article{Jiang:2018pbd,
    author = "Jiang, Minyuan and Craig, Nathaniel and Li, Ying-Ying and Sutherland, Dave",
    title = "{Complete one-loop matching for a singlet scalar in the Standard Model EFT}",
    eprint = "1811.08878",
    archivePrefix = "arXiv",
    primaryClass = "hep-ph",
    doi = "10.1007/JHEP02(2019)031",
    journal = "JHEP",
    volume = "02",
    pages = "031",
    year = "2019",
    note = "[Erratum: JHEP 01, 135 (2021)]"
}

@article{DiLuzio:2017tfn,
    author = {Di Luzio, Luca and Gr\"ober, Ramona and Spannowsky, Michael},
    title = "{Maxi-sizing the trilinear Higgs self-coupling: how large could it be?}",
    eprint = "1704.02311",
    archivePrefix = "arXiv",
    primaryClass = "hep-ph",
    reportNumber = "IPPP-17-26",
    doi = "10.1140/epjc/s10052-017-5361-0",
    journal = "Eur. Phys. J. C",
    volume = "77",
    number = "11",
    pages = "788",
    year = "2017"
}

@article{PhysRevD.16.1519,
  title = {Weak interactions at very high energies: The role of the Higgs-boson mass},
  author = {Lee, Benjamin W. and Quigg, C. and Thacker, H. B.},
  journal = {Phys. Rev. D},
  volume = {16},
  issue = {5},
  pages = {1519--1531},
  numpages = {0},
  year = {1977},
  month = {Sep},
  publisher = {American Physical Society},
  doi = {10.1103/PhysRevD.16.1519},
  url = {https://link.aps.org/doi/10.1103/PhysRevD.16.1519}
}

@article{Englert:2023uug,
    author = "Englert, Christoph and Naskar, Wrishik and Sutherland, Dave",
    title = "{BSM patterns in scalar-sector coupling modifiers}",
    eprint = "2307.14809",
    archivePrefix = "arXiv",
    primaryClass = "hep-ph",
    doi = "10.1007/JHEP11(2023)158",
    journal = "JHEP",
    volume = "11",
    pages = "158",
    year = "2023"
}

@article{Dawson:2023oce,
    author = "Dawson, Sally and Fontes, Duarte and Quezada-Calonge, Carlos and Sanz-Cillero, Juan Jos\'e",
    title = "{Is the HEFT matching unique?}",
    eprint = "2311.16897",
    archivePrefix = "arXiv",
    primaryClass = "hep-ph",
    doi = "10.1103/PhysRevD.109.055037",
    journal = "Phys. Rev. D",
    volume = "109",
    number = "5",
    pages = "055037",
    year = "2024"
}

@article{ATLAS:2023fsi,
    author = "Aad, Georges and others",
    collaboration = "ATLAS",
    title = "{Measurement of the W-boson mass and width with the ATLAS detector using proton{\textendash}proton collisions at $\sqrt{s}=7$ TeV}",
    eprint = "2403.15085",
    archivePrefix = "arXiv",
    primaryClass = "hep-ex",
    reportNumber = "CERN-EP-2024-074",
    doi = "10.1140/epjc/s10052-024-13190-x",
    journal = "Eur. Phys. J. C",
    volume = "84",
    number = "12",
    pages = "1309",
    year = "2024"
}

@article{2024mWCMS,
    author = "Chekhovsky, Vladimir and others",
    collaboration = "CMS",
    title = "{High-precision measurement of the W boson mass with the CMS experiment at the LHC}",
    eprint = "2412.13872",
    archivePrefix = "arXiv",
    primaryClass = "hep-ex",
    reportNumber = "CMS-SMP-23-002, CERN-EP-2024-308",
    month = "12",
    year = "2024"
}

@inproceedings{Papaefstathiou:2022oyi,
    author = "Papaefstathiou, Andreas and Robens, Tania and White, Graham",
    title = "{Signal strength and W-boson mass measurements as a probe of the electro-weak phase transition at colliders - Snowmass White Paper}",
    booktitle = "{Snowmass 2021}",
    eprint = "2205.14379",
    archivePrefix = "arXiv",
    primaryClass = "hep-ph",
    reportNumber = "RBI-ThPhys-2022-20",
    month = "5",
    year = "2022"
}

@article{Dawson:2020oco,
    author = "Dawson, Sally and Homiller, Samuel and Lane, Samuel D.",
    title = "{Putting standard model EFT fits to work}",
    eprint = "2007.01296",
    archivePrefix = "arXiv",
    primaryClass = "hep-ph",
    reportNumber = "YITP-SB-20-18",
    doi = "10.1103/PhysRevD.102.055012",
    journal = "Phys. Rev. D",
    volume = "102",
    number = "5",
    pages = "055012",
    year = "2020"
}

@article{DasBakshi:2024krs,
    author = "Das Bakshi, Supratim and Dawson, Sally and Fontes, Duarte and Homiller, Samuel",
    title = "{Relevance of one-loop SMEFT matching in the 2HDM}",
    eprint = "2401.12279",
    archivePrefix = "arXiv",
    primaryClass = "hep-ph",
    doi = "10.1103/PhysRevD.109.075022",
    journal = "Phys. Rev. D",
    volume = "109",
    number = "7",
    pages = "075022",
    year = "2024"
}

@article{Dawson:2022cmu,
    author = "Dawson, Sally and Fontes, Duarte and Homiller, Samuel and Sullivan, Matthew",
    title = "{Role of dimension-eight operators in an EFT for the 2HDM}",
    eprint = "2205.01561",
    archivePrefix = "arXiv",
    primaryClass = "hep-ph",
    doi = "10.1103/PhysRevD.106.055012",
    journal = "Phys. Rev. D",
    volume = "106",
    number = "5",
    pages = "055012",
    year = "2022"
}

@article{Degrassi:2023eii,
    author = "Degrassi, Giuseppe and Slavich, Pietro",
    title = "{On the two-loop BSM corrections to $h\longrightarrow \gamma \gamma $ in the aligned THDM}",
    eprint = "2307.02476",
    archivePrefix = "arXiv",
    primaryClass = "hep-ph",
    doi = "10.1140/epjc/s10052-023-12097-3",
    journal = "Eur. Phys. J. C",
    volume = "83",
    number = "10",
    pages = "941",
    year = "2023"
}

@article{Branco:2011iw,
    author = "Branco, G. C. and Ferreira, P. M. and Lavoura, L. and Rebelo, M. N. and Sher, Marc and Silva, Joao P.",
    title = "{Theory and phenomenology of two-Higgs-doublet models}",
    eprint = "1106.0034",
    archivePrefix = "arXiv",
    primaryClass = "hep-ph",
    doi = "10.1016/j.physrep.2012.02.002",
    journal = "Phys. Rept.",
    volume = "516",
    pages = "1--102",
    year = "2012"
}

@article{Davidson:2005cw,
    author = "Davidson, Sacha and Haber, Howard E.",
    title = "{Basis-independent methods for the two-Higgs-doublet model}",
    eprint = "hep-ph/0504050",
    archivePrefix = "arXiv",
    reportNumber = "IPPP-03-23, DCPT-03-46, SCIPP-04-15",
    doi = "10.1103/PhysRevD.72.099902",
    journal = "Phys. Rev. D",
    volume = "72",
    pages = "035004",
    year = "2005",
    note = "[Erratum: Phys.Rev.D 72, 099902 (2005)]"
}

@article{Banta:2023prj,
    author = "Banta, Ian and Cohen, Timothy and Craig, Nathaniel and Lu, Xiaochuan and Sutherland, Dave",
    title = "{Effective field theory of the two Higgs doublet model}",
    eprint = "2304.09884",
    archivePrefix = "arXiv",
    primaryClass = "hep-ph",
    reportNumber = "CERN-TH-2023-058",
    doi = "10.1007/JHEP06(2023)150",
    journal = "JHEP",
    volume = "06",
    pages = "150",
    year = "2023"
}

@article{Arco:2020ucn,
    author = "Arco, F. and Heinemeyer, S. and Herrero, M. J.",
    title = "{Exploring sizable triple Higgs couplings in the 2HDM}",
    eprint = "2005.10576",
    archivePrefix = "arXiv",
    primaryClass = "hep-ph",
    reportNumber = "IFT-UAM/CSIC-20-30, FTUAM-20-3",
    doi = "10.1140/epjc/s10052-020-8406-8",
    journal = "Eur. Phys. J. C",
    volume = "80",
    number = "9",
    pages = "884",
    year = "2020"
}

@article{Arco:2025pgx,
    author = {Arco, F. and Heinemeyer, S. and M{\"u}hlleitner, M.},
    title = "{Large one-loop effects of BSM triple Higgs couplings on double Higgs production at $e^+ e^-$ colliders}",
    eprint = "2505.02947",
    archivePrefix = "arXiv",
    primaryClass = "hep-ph",
    reportNumber = "DESY-25-073, IFT--UAM/CSIC-25-016, KA-TP-13-2025",
    doi = "10.1007/JHEP01(2026)160",
    journal = "JHEP",
    volume = "01",
    pages = "160",
    year = "2026"
}

@article{Alloul:2013bka,
    author = "Alloul, Adam and Christensen, Neil D. and Degrande, C{\'e}line and Duhr, Claude and Fuks, Benjamin",
    title = "{FeynRules  2.0 - A complete toolbox for tree-level phenomenology}",
    eprint = "1310.1921",
    archivePrefix = "arXiv",
    primaryClass = "hep-ph",
    reportNumber = "CERN-PH-TH-2013-239, MCNET-13-14, IPPP-13-71, DCPT-13-142, PITT-PACC-1308",
    doi = "10.1016/j.cpc.2014.04.012",
    journal = "Comput. Phys. Commun.",
    volume = "185",
    pages = "2250--2300",
    year = "2014"
}

@article{Crawford:2024nun,
    author = "Crawford, Graeme and Sutherland, Dave",
    title = "{Scalars with non-decoupling phenomenology at future colliders}",
    eprint = "2409.18177",
    archivePrefix = "arXiv",
    primaryClass = "hep-ph",
    doi = "10.1007/JHEP04(2025)197",
    journal = "JHEP",
    volume = "04",
    pages = "197",
    year = "2025"
}

@article{Brivio:2020onw,
    author = "Brivio, Ilaria",
    title = "{SMEFTsim 3.0 {\textemdash} a practical guide}",
    eprint = "2012.11343",
    archivePrefix = "arXiv",
    primaryClass = "hep-ph",
    doi = "10.1007/JHEP04(2021)073",
    journal = "JHEP",
    volume = "04",
    pages = "073",
    year = "2021"
}

@article{Heinrich:2022gzl,
    author = "Heinrich, Gudrun and Lang, Jannis and Scyboz, Ludovic",
    title = "{Beyond dimension six in SM Effective Field Theory: a case study in Higgs pair production at NLO QCD}",
    eprint = "2207.08790",
    archivePrefix = "arXiv",
    primaryClass = "hep-ph",
    reportNumber = "KA-TP-21-2022, P3H-22-079",
    doi = "10.22323/1.416.0009",
    journal = "PoS",
    volume = "LL2022",
    pages = "009",
    year = "2022"
}

@article{Dittmaier:2021fls,
    author = "Dittmaier, Stefan and Schuhmacher, Sebastian and Stahlhofen, Maximilian",
    title = "{Integrating out heavy fields in the path integral using the background-field method: general formalism}",
    eprint = "2102.12020",
    archivePrefix = "arXiv",
    primaryClass = "hep-ph",
    reportNumber = "FR-PHENO-2020-010",
    doi = "10.1140/epjc/s10052-021-09587-7",
    journal = "Eur. Phys. J. C",
    volume = "81",
    number = "9",
    pages = "826",
    year = "2021"
}

@article{ATLAS:2024lyh,
    author = "Aad, Georges and others",
    collaboration = "ATLAS",
    title = "{Interpretations of the ATLAS measurements of Higgs boson production and decay rates and differential cross-sections in pp collisions at $ \sqrt{s} $ = 13 TeV}",
    eprint = "2402.05742",
    archivePrefix = "arXiv",
    primaryClass = "hep-ex",
    reportNumber = "CERN-EP-2024-017",
    doi = "10.1007/JHEP11(2024)097",
    journal = "JHEP",
    volume = "11",
    pages = "097",
    year = "2024"
}

@article{Degrassi:2025pqt,
    author = {Degrassi, Giuseppe and Gr{\"o}ber, Ramona and Slavich, Pietro},
    title = "{Two-loop BSM contributions to Higgs pair production in the aligned THDM}",
    eprint = "2508.11539",
    archivePrefix = "arXiv",
    primaryClass = "hep-ph",
    doi = "10.1007/JHEP01(2026)041",
    journal = "JHEP",
    volume = "01",
    pages = "041",
    year = "2026"
}

@article{Kanemura:2002vm,
    author = "Kanemura, Shinya and Kiyoura, Shingo and Okada, Yasuhiro and Senaha, Eibun and Yuan, C. P.",
    title = "{New physics effect on the Higgs selfcoupling}",
    eprint = "hep-ph/0211308",
    archivePrefix = "arXiv",
    reportNumber = "KEK-TH-856, MSUHEP-21119",
    doi = "10.1016/S0370-2693(03)00268-5",
    journal = "Phys. Lett. B",
    volume = "558",
    pages = "157--164",
    year = "2003"
}

@article{Braathen:2019pxr,
    author = "Braathen, Johannes and Kanemura, Shinya",
    title = "{On two-loop corrections to the Higgs trilinear coupling in models with extended scalar sectors}",
    eprint = "1903.05417",
    archivePrefix = "arXiv",
    primaryClass = "hep-ph",
    reportNumber = "OU-HET-1001",
    doi = "10.1016/j.physletb.2019.07.021",
    journal = "Phys. Lett. B",
    volume = "796",
    pages = "38--46",
    year = "2019"
}

@article{Braathen:2019zoh,
    author = "Braathen, Johannes and Kanemura, Shinya",
    title = "{Leading two-loop corrections to the Higgs boson self-couplings in models with extended scalar sectors}",
    eprint = "1911.11507",
    archivePrefix = "arXiv",
    primaryClass = "hep-ph",
    reportNumber = "OU-HET-1030",
    doi = "10.1140/epjc/s10052-020-7723-2",
    journal = "Eur. Phys. J. C",
    volume = "80",
    number = "3",
    pages = "227",
    year = "2020"
}

@article{Kanemura:2004mg,
    author = "Kanemura, Shinya and Okada, Yasuhiro and Senaha, Eibun and Yuan, C. -P.",
    title = "{Higgs coupling constants as a probe of new physics}",
    eprint = "hep-ph/0408364",
    archivePrefix = "arXiv",
    doi = "10.1103/PhysRevD.70.115002",
    journal = "Phys. Rev. D",
    volume = "70",
    pages = "115002",
    year = "2004"
}

@article{Bahl:2023eau,
    author = "Bahl, Henning and Braathen, Johannes and Gabelmann, Martin and Weiglein, Georg",
    title = "{anyH3: precise predictions for the trilinear Higgs coupling in the Standard Model and beyond}",
    eprint = "2305.03015",
    archivePrefix = "arXiv",
    primaryClass = "hep-ph",
    doi = "10.1140/epjc/s10052-023-12173-8",
    journal = "Eur. Phys. J. C",
    volume = "83",
    number = "12",
    pages = "1156",
    year = "2023",
    note = "[Erratum: Eur.Phys.J.C 84, 498 (2024)]"
}

@article{Brivio:2025sib,
    author = {Brivio, Ilaria and Gr{\"o}ber, Ramona and Schmid, Konstantin},
    title = "{Higgs pair production in gluon fusion to higher orders in Higgs Effective Field Theory}",
    eprint = "2511.23411",
    archivePrefix = "arXiv",
    primaryClass = "hep-ph",
    reportNumber = "COMETA-2025-52",
    month = "11",
    year = "2025"
}

@article{Brivio:2025yrr,
    author = {Brivio, Ilaria and Gr{\"o}ber, Ramona and Schmid, Konstantin},
    title = "{The Art of Counting: a reappraisal of the HEFT expansion}",
    eprint = "2511.23410",
    archivePrefix = "arXiv",
    primaryClass = "hep-ph",
    reportNumber = "COMETA-2025-51",
    month = "11",
    year = "2025"
}

@article{Gavela:2016bzc,
    author = "Gavela, B. M. and Jenkins, E. E. and Manohar, A. V. and Merlo, L.",
    title = "{Analysis of General Power Counting Rules in Effective Field Theory}",
    eprint = "1601.07551",
    archivePrefix = "arXiv",
    primaryClass = "hep-ph",
    reportNumber = "CERN-TH-2016-015, FTUAM-16-2, IFT-UAM-CSIC-16-006",
    doi = "10.1140/epjc/s10052-016-4332-1",
    journal = "Eur. Phys. J. C",
    volume = "76",
    number = "9",
    pages = "485",
    year = "2016"
}

@article{Buchalla:2013eza,
    author = "Buchalla, Gerhard and Cat\'a, Oscar and Krause, Claudius",
    title = "{On the Power Counting in Effective Field Theories}",
    eprint = "1312.5624",
    archivePrefix = "arXiv",
    primaryClass = "hep-ph",
    reportNumber = "LMU-ASC-81-13",
    doi = "10.1016/j.physletb.2014.02.015",
    journal = "Phys. Lett. B",
    volume = "731",
    pages = "80--86",
    year = "2014"
}

@article{Grober:2025vse,
    author = {Gr{\"o}ber, Ramona and Rossia, Alejo N. and Ryczkowski, Micha{\l}},
    title = "{Multi-Higgs Amplitudes Bootstrapped: Dissecting SMEFT and HEFT}",
    eprint = "2509.02680",
    archivePrefix = "arXiv",
    primaryClass = "hep-ph",
    reportNumber = "COMETA-2025-34",
    month = "9",
    year = "2025"
}

@article{Bagnaschi:2023rbx,
    author = {Bagnaschi, Emanuele and Degrassi, Giuseppe and Gr{\"o}ber, Ramona},
    title = "{Higgs boson pair production at NLO in the POWHEG approach and the top quark mass uncertainties}",
    eprint = "2309.10525",
    archivePrefix = "arXiv",
    primaryClass = "hep-ph",
    reportNumber = "CERN-TH-2023-131",
    doi = "10.1140/epjc/s10052-023-12238-8",
    journal = "Eur. Phys. J. C",
    volume = "83",
    number = "11",
    pages = "1054",
    year = "2023"
}

@article{Buchalla:2016bse,
    author = "Buchalla, G. and Cata, O. and Celis, A. and Krause, C.",
    title = "{Standard Model Extended by a Heavy Singlet: Linear vs. Nonlinear EFT}",
    eprint = "1608.03564",
    archivePrefix = "arXiv",
    primaryClass = "hep-ph",
    reportNumber = "LMU-ASC-35-16",
    doi = "10.1016/j.nuclphysb.2017.02.006",
    journal = "Nucl. Phys. B",
    volume = "917",
    pages = "209--233",
    year = "2017"
}

@article{Heinrich:2023rsd,
    author = "Heinrich, Gudrun and Lang, Jannis",
    title = "{Combining chromomagnetic and four-fermion operators with leading SMEFT operators for $gg \to  hh$ at NLO QCD}",
    eprint = "2311.15004",
    archivePrefix = "arXiv",
    primaryClass = "hep-ph",
    reportNumber = "KA-TP-29-2023,P3H-23-095, KA-TP-29-2023, P3H-23-095",
    doi = "10.1007/JHEP05(2024)121",
    journal = "JHEP",
    volume = "05",
    pages = "121",
    year = "2024"
}

@article{Grober:2010yv,
    author = "Grober, R. and Muhlleitner, M.",
    title = "{Composite Higgs Boson Pair Production at the LHC}",
    eprint = "1012.1562",
    archivePrefix = "arXiv",
    primaryClass = "hep-ph",
    reportNumber = "KA-TP-37-2010",
    doi = "10.1007/JHEP06(2011)020",
    journal = "JHEP",
    volume = "06",
    pages = "020",
    year = "2011"
}

@article{Grober:2016wmf,
    author = "Grober, Ramona and Muhlleitner, Margarete and Spira, Michael",
    title = "{Signs of Composite Higgs Pair Production at Next-to-Leading Order}",
    eprint = "1602.05851",
    archivePrefix = "arXiv",
    primaryClass = "hep-ph",
    reportNumber = "KA-TP-05-2016, PSI-PR-16-02, RM3-TH-16-3",
    doi = "10.1007/JHEP06(2016)080",
    journal = "JHEP",
    volume = "06",
    pages = "080",
    year = "2016"
}

@article{Contino:2012xk,
    author = "Contino, Roberto and Ghezzi, Margherita and Moretti, Mauro and Panico, Giuliano and Piccinini, Fulvio and Wulzer, Andrea",
    title = "{Anomalous Couplings in Double Higgs Production}",
    eprint = "1205.5444",
    archivePrefix = "arXiv",
    primaryClass = "hep-ph",
    doi = "10.1007/JHEP08(2012)154",
    journal = "JHEP",
    volume = "08",
    pages = "154",
    year = "2012"
}

@article{Dedes:2025oda,
    author = "Dedes, Athanasios and Rosiek, Janusz and Ryczkowski, Micha{\l}",
    title = "{Double Higgs boson production via vector boson fusion in SMEFT}",
    eprint = "2506.12917",
    archivePrefix = "arXiv",
    primaryClass = "hep-ph",
    reportNumber = "COMETA-2025-33",
    doi = "10.1103/bywd-2dy1",
    journal = "Phys. Rev. D",
    volume = "112",
    number = "5",
    pages = "055044",
    year = "2025"
}

@misc{ggHHdim8,
    author = "Brivio, Ilaria and Gr{\"o}ber, Ramona and Mimasu, Ken and Schmid, Konstantin",
    howpublished = "To appear",
    title = "Di-Higgs production from gluon fusion at dimension-8 in SMEFT"
}

@article{Grober:2017gut,
    author = "Grober, R. and Muhlleitner, M. and Spira, M.",
    title = "{Higgs Pair Production at NLO QCD for CP-violating Higgs Sectors}",
    eprint = "1705.05314",
    archivePrefix = "arXiv",
    primaryClass = "hep-ph",
    doi = "10.1016/j.nuclphysb.2017.10.002",
    journal = "Nucl. Phys. B",
    volume = "925",
    pages = "1--27",
    year = "2017"
}

@misc{hpair,
  author       = {{Michael Spira}},
  title        = {hpair},
  howpublished = {\url{https://https://ltpth.pages.psi.ch/tiger/}},
  note         = {Accessed: 2026-01-15}
}

@article{Grober:2015cwa,
    author = "Grober, Ramona and Muhlleitner, Margarete and Spira, Michael and Streicher, Juraj",
    title = "{NLO QCD Corrections to Higgs Pair Production including Dimension-6 Operators}",
    eprint = "1504.06577",
    archivePrefix = "arXiv",
    primaryClass = "hep-ph",
    reportNumber = "KA-TP-08-2015, PSI-PR-15-03, RM3-TH-15-5",
    doi = "10.1007/JHEP09(2015)092",
    journal = "JHEP",
    volume = "09",
    pages = "092",
    year = "2015"
}

@article{Kilic:2026ogm,
    author = "Kilic, Can and Mathai, Sanjay and Youn, Taewook",
    title = "{Constraints on Loryons in a Two Higgs Doublet Model}",
    eprint = "2601.14389",
    archivePrefix = "arXiv",
    primaryClass = "hep-ph",
    reportNumber = "UT-WI-02-2026",
    month = "1",
    year = "2026"
}

@article{Herrero:2022krh,
    author = "Herrero, M. J. and Morales, R. A.",
    title = "{One-loop corrections for WW to HH in Higgs EFT with the electroweak chiral Lagrangian}",
    eprint = "2208.05900",
    archivePrefix = "arXiv",
    primaryClass = "hep-ph",
    reportNumber = "IFT-UAM/CSIC-22-80",
    doi = "10.1103/PhysRevD.106.073008",
    journal = "Phys. Rev. D",
    volume = "106",
    number = "7",
    pages = "073008",
    year = "2022"
}

@article{Arco:2023sac,
    author = "Arco, F. and Domenech, D. and Herrero, M. J. and Morales, R. A.",
    title = "{Nondecoupling effects from heavy Higgs bosons by matching 2HDM to HEFT amplitudes}",
    eprint = "2307.15693",
    archivePrefix = "arXiv",
    primaryClass = "hep-ph",
    reportNumber = "IFT-UAM/CSIC-23-97",
    doi = "10.1103/PhysRevD.108.095013",
    journal = "Phys. Rev. D",
    volume = "108",
    number = "9",
    pages = "095013",
    year = "2023"
}

@article{Buchalla:2018yce,
    author = "Buchalla, G. and Capozi, M. and Celis, A. and Heinrich, G. and Scyboz, L.",
    title = "{Higgs boson pair production in non-linear Effective Field Theory with full $m_t$-dependence at NLO QCD}",
    eprint = "1806.05162",
    archivePrefix = "arXiv",
    primaryClass = "hep-ph",
    reportNumber = "LMU-ASC 34/18, MPP-2018-127, LMU-ASC-34-18",
    doi = "10.1007/JHEP09(2018)057",
    journal = "JHEP",
    volume = "09",
    pages = "057",
    year = "2018",
    note = "[Erratum: JHEP 06, 094 (2025)]"
}

@article{Heinrich:2022idm,
    author = "Heinrich, Gudrun and Lang, Jannis and Scyboz, Ludovic",
    title = "{SMEFT predictions for $gg \to hh$ at full NLO QCD and truncation uncertainties}",
    eprint = "2204.13045",
    archivePrefix = "arXiv",
    primaryClass = "hep-ph",
    reportNumber = "KA-TP-14-2022, OUTP-22-05P, P3H-22-045",
    doi = "10.1007/JHEP08(2022)079",
    journal = "JHEP",
    volume = "08",
    pages = "079",
    year = "2022",
    note = "[Erratum: JHEP 10, 086 (2023)]"
}

@article{Djouadi:2018xqq,
    author = "Djouadi, Abdelhak and Kalinowski, Jan and Muehlleitner, Margarete and Spira, Michael",
    collaboration = "HDECAY",
    title = "{HDECAY: Twenty$_{++}$ years after}",
    eprint = "1801.09506",
    archivePrefix = "arXiv",
    primaryClass = "hep-ph",
    reportNumber = "LPT-ORSAY-18-04, CERN-TH-2017-262, LPT-Orsay-18-04, KA-TP-03-2018, PSI-PR-18-02",
    doi = "10.1016/j.cpc.2018.12.010",
    journal = "Comput. Phys. Commun.",
    volume = "238",
    pages = "214--231",
    year = "2019"
}

@article{Djouadi:1997yw,
    author = "Djouadi, A. and Kalinowski, J. and Spira, M.",
    title = "{HDECAY: A Program for Higgs boson decays in the standard model and its supersymmetric extension}",
    eprint = "hep-ph/9704448",
    archivePrefix = "arXiv",
    reportNumber = "DESY-97-079, IFT-96-29, PM-97-04",
    doi = "10.1016/S0010-4655(97)00123-9",
    journal = "Comput. Phys. Commun.",
    volume = "108",
    pages = "56--74",
    year = "1998"
}

@article{Braun:2025hvr,
    author = {Braun, Jens and Bredt, Pia and Heinrich, Gudrun and H{\"o}fer, Marius},
    title = "{Double Higgs production in vector boson fusion at NLO QCD in HEFT}",
    eprint = "2502.09132",
    archivePrefix = "arXiv",
    primaryClass = "hep-ph",
    reportNumber = "KA-TP-03-2025, SI-HEP-2025-03, P3H-25-010",
    doi = "10.1007/JHEP07(2025)209",
    journal = "JHEP",
    volume = "07",
    pages = "209",
    year = "2025"
}

@article{Jager:2025isz,
    author = {J{\"a}ger, Barbara and Karlberg, Alexander and Reinhardt, Simon},
    title = "{Precision tools for the simulation of double-Higgs production via vector-boson fusion}",
    eprint = "2502.09112",
    archivePrefix = "arXiv",
    primaryClass = "hep-ph",
    reportNumber = "CERN-TH-2025-030",
    doi = "10.1007/JHEP06(2025)022",
    journal = "JHEP",
    volume = "06",
    pages = "022",
    year = "2025"
}

@article{DiMicco:2019ngk,
    author = "Alison, J. and others",
    editor = "Di Micco, Biagio and Gouzevitch, Maxime and Mazzitelli, Javier and Vernieri, Caterina",
    title = "{Higgs boson potential at colliders: Status and perspectives}",
    eprint = "1910.00012",
    archivePrefix = "arXiv",
    primaryClass = "hep-ph",
    reportNumber = "FERMILAB-CONF-19-468-E-T, LHCXSWG-2019-005",
    doi = "10.1016/j.revip.2020.100045",
    journal = "Rev. Phys.",
    volume = "5",
    pages = "100045",
    year = "2020"
}

@article{Buchalla:2022vjp,
    author = {Buchalla, Gerhard and Heinrich, Gudrun and M{\"u}ller-Salditt, Ch. and Pandler, Florian},
    title = "{Loop counting matters in SMEFT}",
    eprint = "2204.11808",
    archivePrefix = "arXiv",
    primaryClass = "hep-ph",
    reportNumber = "LMU-ASC{\textasciitilde}16/22, KA-TP-10-2022, P3H-22-041",
    doi = "10.21468/SciPostPhys.15.3.088",
    journal = "SciPost Phys.",
    volume = "15",
    number = "3",
    pages = "088",
    year = "2023"
}

@article{Ge:2026qfa,
    author = "Ge, Zizhou and Song, Huayang and Wan, Xia",
    title = "{Establishing the Primary HEFT as a Precision Benchmark for UV-HEFT Matching}",
    eprint = "2602.14418",
    archivePrefix = "arXiv",
    primaryClass = "hep-ph",
    month = "2",
    year = "2026"
}

@article{Song:2025kjp,
    author = "Song, Huayang and Wan, Xia",
    title = "{Matching the real Higgs triplet extension of Standard Model to HEFT}",
    eprint = "2503.00707",
    archivePrefix = "arXiv",
    primaryClass = "hep-ph",
    doi = "10.1007/JHEP06(2025)249",
    journal = "JHEP",
    volume = "06",
    pages = "249",
    year = "2025"
}
%%%%%%%%%%%%%%%%%%%

\end{document}